\documentclass[a4paper]{article}%
\usepackage[utf8]{inputenc}
\usepackage[T1]{fontenc}
\usepackage{lmodern}
\usepackage[ngerman,english]{babel}
\usepackage{a4wide}
\usepackage[affil-it]{authblk}
\usepackage[automark]{scrpage2}
\usepackage{graphicx}
\usepackage{blindtext}
\usepackage{cancel}
\usepackage{amsmath}
\usepackage{amssymb}
\usepackage{dsfont}
\usepackage{tensor}
\usepackage{slashed}
\usepackage{multirow}
\usepackage[colorlinks,pdfpagelabels, pdfstartview = FitH, bookmarksopen =
true, bookmarksnumbered = true, linkcolor = blue, plainpages = false,
hypertexnames = false, citecolor = blue]{hyperref}
\usepackage[toc,page]{appendix}
\usepackage{cite}
\usepackage{amsfonts}%
\numberwithin{equation}{section}

\begin{document}

\title{Gauge Theory by canonical Transformations}
\author{Adrian Koenigstein\textsuperscript{1,2}, 
Johannes Kirsch\textsuperscript{2}, Horst Stoecker\textsuperscript{1,2,3}, Juergen Struckmeier\textsuperscript{1,2,3}, David Vasak\textsuperscript{2}, Matthias Hanauske\textsuperscript{1,2}}
\affil{\textsuperscript{1}Institut f\"ur Theoretische Physik, Johann Wolfgang 
Goethe-Universit\"at\\ Max-von-Laue-Str.\ 1, 60438 Frankfurt am Main, Germany}
\affil{\textsuperscript{2}Frankfurt Institute for Advanced Studies\\ Ruth-Moufang-Str.\ 1, 
60438 Frankfurt am Main, Germany}
\affil{\textsuperscript{3}Helmholtzzentrum f\"ur Schwerionenforschung (GSI)\\ Planckstraße\ 1, 64291 Darmstadt, Germany}
\maketitle
\thispagestyle{empty}

\begin{abstract}
Electromagnetism, the strong and the weak interaction are commonly formulated as gauge theories in a Lagrangian description. In this paper we present an alternative formal derivation of $U(1)$-gauge theory in a manifestly covariant Hamilton formalism. We make use of canonical transformations as our guiding tool to formalize the gauging procedure. The introduction of the gauge field, its transformation behaviour and a dynamical gauge field Lagrangian/Hamiltonian are unavoidable consequences of this formalism, whereas the form of the free gauge Lagrangian/Hamiltonian depends on the selection of the gauge dependence of the canonically conjugate gauge fields.
\end{abstract}

\tableofcontents
\setcounter{tocdepth}{2}

\setcounter{page}{2}

\section{Introduction}
\label{sec:introduction}
Except for the gravitational interaction, all known forces (electromagnetism, weak and strong nuclear interaction) are successfully deduced from gauge theories \cite{fock26,weyl29,tHooft94}. The concept of the invariance of the Lagrangian $\mathcal{L}$ under local symmetry transformations generates additional gauge fields which induce the interaction of the underlying symmetry group. The main objective of the present article is to formalize the gauge theoretical concept in a covariant Hamilton formulation \cite{dedonder30,weyl35,musi78,guenther87,tapia88,sarda95,paufler01,struckmeier08}, applied to $U(1)$ phase transformations:
\\
We start with a short repetition and summary of conventional Lagrangian formulation of gauge theory, exemplified for scalar electrodynamics (Chap. \ref{sec:conventionalgaugetheory}).
\\
In Chap.\ \ref{sec:covarianthamiltondensity} a covariant canonical Hamilton field theory is introduced, formulated and derived from conventional Lagrangian field theory, thereby ensuring that the principle of least action is maintained as the fundamental guiding principle. The general formalism of covariant canonical transformations is discussed (Sec. \ref{sec:canonicaltransformations}), compare \cite{good54,musi78,struckmeier08}.
\\
The second part of the paper exemplifies the procedure of this new canonical strategy for the simplest non-trivial case, i.e. scalar electrodynamics ($U(1)$-symmetry), see Chap. \ref{sec:scalarelectrodynamics}. The local phase transformations are presented as canonical transformations (Sec. \ref{sec:localphasetransformations}) and the corresponding gauge field dynamics follows directly (Sec.\ \ref{sec:gaugefielddynamics}). Thereby the formalism suggests minimal coupling, as the simplest electro-magnetic interaction, which in turn enforces the introduction of a kinetic Hamiltonian/Lagrangian for the gauge fields. For a specific choice of gauge of canonically conjugate fields, Maxwell equations as well as the Maxwell Lagrangian (Sec. \ref{sec:kgmaxwellhamilton} and \ref{sec:kgmaxwelllagrange}) are derived.\\
In the outlook and during the elaboration we discuss open questions especially remaining arbitrariness in choosing integration constants in the generating function of the canonical transformation, which might lead to different dynamical Hamiltonians/Lagrangians for the gauge field. We sketch and refer on ideas for generalisations to higher symmetry groups including $SU(N)$-symmetry and other massive matter fields in the outlook.
\\
\\
Our units are $\hbar = c =1$ and the metric tensor is $\eta_{\mu\nu} = \mathrm{diag}(+,-,-,-)$. Space-time dependency $x^\mu$ of a quantity is abbreviated $x$.

\section{Conventional formulation of $U(1)$-gauge theory}
\label{sec:conventionalgaugetheory}

In the following the main concepts of gauge theory will briefly be illustrated and repeated by using the example of scalar electrodynamics (for details see e.g.\ \cite{cheng1984gauge,dehnen1999theorie,hanauske2005eigenschaften}): 
\\
\\
Starting from the globally $U(1)$-invariant Lagrangian of a complex scalar field $\phi$ without spin
	\begin{align}
		\mathcal{L} & = (\partial_\mu \bar{\phi}) (\partial^\mu \phi) - m^2 \bar{\phi}\, \phi \quad \,, \label{eq:lagr1}
	\end{align}
the concept of gauge theory demands that this Lagrangian should be invariant under local unitary $U(1)$ phase transformations
	\begin{align}
		\phi \longrightarrow \Phi & = e^{-i \Lambda(x)}\, \phi \vphantom{\overset{!}{=}} \,,
		\\
		\bar{\phi} \longrightarrow \bar{\Phi} & = \bar{\phi} \, e^{i \Lambda(x)} \vphantom{\overset{!}{=}} \,,
		\\
		\mathcal{L} (\Phi,\, \bar{\Phi},\, \partial_\alpha \Phi,\, \partial_\alpha \bar{\Phi},\, x) & \overset{!}{=} \mathcal{L} (\phi,\, \bar{\phi},\, \partial_\alpha \phi,\, \partial_\alpha \bar{\phi},\, x) \,. \label{eq:request}
	\end{align}
Since the group parameter $\Lambda(x)$ depends on space and time, the first term of the Lagrangian (\ref{eq:lagr1}) gains additional terms after phase transformation (we drop the $x$-dependency in the following). In order to maintain the invariance of the Lagrangian and compensate the additional terms, a gauge field $a_\mu$ has to be implemented into the theory. Gauge theory incorporates these auxiliary interactions using the concept of covariant derivatives and minimal coupling. It can be shown that by converting the partial derivatives into covariant derivatives 
	\begin{align}
		\partial_\mu \longrightarrow \mathcal{D}_\mu & \equiv \partial_\mu - i q \, a_\mu  
	\end{align}
the request (\ref{eq:request}) is achieved if the gauge potentials obeys the following inhomogeneous transformation rule
	\begin{align}
		A_\mu & = a_\mu + \frac{1}{q} (\partial_\mu \Lambda) \,.
	\end{align}
By using the covariant derivative, the Lagrangian (\ref{eq:lagr1}) of the free particle is converted into a Lagrangian where the complex (charged) scalar field interacts electromagnetically. However, to close the system, the kinetic terms of the gauge potentials $a_\mu$ need to be added by hand to the Lagrangian in a covariant way. In gauge theory, the contracted quadratic term of the gauge field strength tensor $F_{\mu\nu}$,
	\begin{align}
		F_{\mu\nu} & \equiv \frac{1}{iq} \left[ \mathcal{D}_\mu , \mathcal{D}_\nu \right]_{-} = \partial_\mu a_\nu - \partial_\nu a_\mu + iq \underbrace{\left[ a_\mu , a_\nu \right]_{-}}_{= 0\,\,\text{for U(1)}}
	\end{align}
is usually added to the Lagrangian
	\begin{align}
		\mathcal{L} (a_\alpha,\, \partial_\beta a_\alpha) & = - \frac{1}{4}  F_{\mu \nu} F^{\mu \nu} \,,
	\end{align}
where upon this choice is not unique and terms of higher order could be added arbitrary. In conclusion, the final form of the Lagrangian, which is invariant under local $U(1)$ gauge transformations, has the following form
	\begin{align}
		\mathcal{L} (\phi,\, \bar{\phi},\, \partial_\beta \phi,\, \partial_\beta \bar{\phi},\, a_\alpha,\, \partial_\beta a_\alpha) & = (\mathcal{D}^{*}_\mu \bar{\phi}) (\mathcal{D}^\mu \phi) - m^2 \bar{\phi}\, \phi - \frac{1}{4} F_{\mu\nu} F^{\mu\nu} \,.
	\end{align}

\section{Covariant canonical field theory}
\label{sec:covarianthamiltondensity}
In this section we give a short repetition of the Lagrange formalism in classical field theory, compare \cite{greiner96,saletan98}, and introduce the covariant Hamilton formalism for classical fields, which is known in literature as DeDonder-Weyl formalism \cite{dedonder30,weyl35}. We will focus on the main aspects here and refer the reader to \cite{struckmeier08} for more details. For the sake of simplicity we restrict ourself to a real scalar field in the first chapter. A generalisation to complex scalar fields will be applied in the next chapter \ref{sec:scalarelectrodynamics}.

\subsection{Lagrangian field theory}
\label{sec:lagrangianfieldtheory}
The Lagrangian description of field theories is based on the Lagrange density $\mathcal{L}$ which is supposed to contain all information about a given physical system. In a first order field theory the Lagrange density $\mathcal{L}$ only depends on the physical fields $\phi(x)$ and their derivatives $\partial_\mu \phi(x)$, which are mutually independent. In addition, $\mathcal{L}$ may depend on the space-time position $x$ explicitly,
	\begin{align}
		\mathcal{L} & \equiv \mathcal{L}(\phi,\, \partial_\alpha \phi,\, x) \,.
	\end{align}
The variational principle yields the Euler-Lagrange equation, see App.\ \ref{app:euler-lagrangeequation}, describing the dynamics of the system,
	\begin{align}
		\partial_\mu \frac{\partial \mathcal{L}}{\partial (\partial_\mu \phi)} - \frac{\partial \mathcal{L}}{\partial \phi} & = 0 \,. \label{eq:eulerlagrangeequation}
	\end{align}
However, a variety of Lagrange densities can lead to the same Euler-Lagrange equation (\ref{eq:eulerlagrangeequation}) and hence describe identical physical systems:
\\
\\
1. Obviously this is the case if Lagrange densities differ by total divergences of vector functions $\partial_\mu \mathcal{F}^\mu (\phi,\, x)$, which can depend on $\phi$ and $x$ only. In consequence
	\begin{align}
		\mathcal{L^\prime}(\phi,\, \partial_\alpha \phi,\, x) & = \mathcal{L}(\phi,\, \partial_\alpha \phi,\, x) + \partial_\mu \mathcal{F}^\mu (\phi,\, x) \label{eq:surfaceterms} \,.
	\end{align}
This relation will be important for the derivation of canonical transformations, see Sec.\ \ref{sec:canonicaltransformations}. These total divergences of vector functions $\partial_\mu \mathcal{F}^\mu (\phi,\, x)$ are usually referred to as surface terms in field theory\footnote{In point mechanics they are called mechanical gauge transformations.}. We will show in App.\ \ref{app:surfaceterms} that they can be neglected or added to a Lagrange density (for appropriate boundary conditions) without changing the dynamics\footnote{Notice that by adding a term that explicitly depends on the spacetime point $x$ the observables derived from space-time-symmetries (via the Noether theorem) are modified. This is in particular the energy-momentum tensor  and angular momentum tensor linked to space-time homogeneity and isotropy. If energy is not conserved this corresponds to ``opening up'' a closed system. However, as canonical transformations (see Sec.\ \ref{sec:canonicaltransformations})) are reversible this formalism allows for selecting open systems where the solution of the equations of motion is particularly simple, and then transforming that solution back into the original closed system. In the following (see Sec.\ \ref{sec:scalarelectrodynamics}) we will however take a different path, namely closing the modified open system emerging from the requirement of form invariance of the Hamiltonian by adding degrees of freedom (gauge fields) compatible with the required system symmetry that account for the ``external'' energy and momenta.}.
\\
\\
2. While the choice of generalised coordinates (in field theory the choice of field representations) is not fixed, the total number of generalized coordinates (fields) is and determines the degrees of freedom in the system. From point mechanics we know, that all possible choices of generalised coordinates are linked by so called point transformations. The same applies for field theory. If the map $\Phi \rightarrow \phi$, and respectively $\phi \rightarrow \Phi$, is differentiable
	\begin{align}
		\phi & = \phi (\Phi,\, x) \,,
		\\
		\Phi & = \Phi (\phi,\, x)		
	\end{align}
and $\mathcal{L}(\phi,\, \partial_\alpha \phi,\, x)$ satisfies the Euler-Lagrange equation (\ref{eq:eulerlagrangeequation}), then
	\begin{align}
		\tilde{\mathcal{L}}\left( \Phi,\, \partial_\beta \Phi,\, x \right) & = \mathcal{L} \left[\phi(\Phi,\, x),\, \partial_\alpha \phi(\Phi,\, \partial_\beta \Phi,\, x),\, x \right] \label{eq:pointtransformationsforfields}
	\end{align}
maintains the form of Euler-Lagrange equation
	\begin{align}
		\partial_\mu \frac{\partial \tilde{\mathcal{L}}}{\partial (\partial_\mu \Phi)} - \frac{\partial \tilde{\mathcal{L}}}{\partial \Phi} & = 0 \,,
	\end{align}
as well vice versa. This means that $\tilde{\mathcal{L}}$ and $\mathcal{L}$ both describe the identical system and same physics; for a proof see App.\ \ref{app:pointtransformations}.
\\
\\
Combining invariance under additional surface terms (1.) and under point transformations (2.), one finds with the respective Eqs.\ (\ref{eq:surfaceterms}) and (\ref{eq:pointtransformationsforfields}) that
	\begin{align}
		\tilde{\mathcal{L}}^\prime \left( \Phi,\, \partial_\beta \Phi,\, x \right) & = \tilde{\mathcal{L}} \left( \Phi,\, \partial_\beta \Phi,\, x \right) + \partial_\mu F^\mu(\Phi,\, x) \nonumber
		\\
		& = \mathcal{L}^\prime \left[\phi(\Phi,\, x),\, \partial_\alpha \phi(\Phi,\, \partial_\beta \Phi,\, x),\, x \right] + \partial_\mu F^\mu(\Phi,\, x) \nonumber
		\\
		& = \mathcal{L}^\prime (\phi,\, \partial_\alpha \phi,\, x) + \partial_\mu F^\mu(\Phi,\, x) \nonumber
		\\
		& = \mathcal{L}(\phi,\, \partial_\alpha \phi,\, x) + \partial_\mu f^\mu_1 \left( \phi,\, \Phi,\, x \right) \,, \label{eq:lagrangians}
	\end{align}
which will be the starting point of Sec.\ \ref{sec:canonicaltransformations}, where canonical transformations will be discussed.\footnote{At this point we remark that also the multiplication of the Lagrangian with a global scale factor $\lambda$ leaves the Euler-Lagrange equations (\ref{eq:eulerlagrangeequation}) invariant and hence does not change the dynamics of the system. This might be obvious in the Lagrangian description, but leads to difficulties in the context of canonical transformations, since for example phase space volume is not conserved. Therefore we will exclude this kind of transformations later on, see Sec.\ \ref{sec:canonicaltransformations}.}

\subsection{Covariant Hamilton field theory}
\label{eq:covarianthamiltonfieldtheory}
In point mechanics the Hamilton formulation is not simply a completely equivalent description to the Newtonian and Lagrangian theory \cite{saletan98,greiner03}. It has the advantage to allow deploying the powerful method of covariant canonical transformations.
\\
Notice that conventional Hamilton field theories do not have this feature, since they are not manifestly covariant as time and space derivatives of fields are treated in different ways, compare \cite{weinberg96,greiner96,saletan98}. The conventional Hamilton density $H$ depends on the field $\phi$, the field $\pi$ that is canonically conjugate to $\phi$ and the spatial derivative $\vec{\nabla} \phi$,
	\begin{align}
		H & \equiv H(\phi, \pi, \vec{\nabla}\phi, x) \,.
	\end{align}
In contrast to this conventional description, where the canonically conjugate field $\pi$ corresponds only to the derivative with respect to time of the field $\phi$,
	\begin{align}
		\pi & \equiv \frac{\partial \mathcal{L}}{\partial (\partial_0 \phi)} \,,
	\end{align}
we develop an extended Hamilton formalism for field theory based on a manifestly covariant Legendre transformation. In this so called DeDonder-Weyl approach, see \cite{dedonder30,weyl35}, the canonically conjugate field $\pi^\mu$ is defined as follows,
 	\begin{align}
	 	\pi^\mu & \equiv \frac{\partial \mathcal{L}}{\partial (\partial_\mu \phi)} \,. \label{eq:canonicallyconjugatefield}
 	\end{align}
Now the covariant Legendre transformation
	\begin{align}
		\mathcal{H}(\phi, \pi^\alpha, x) & \equiv \pi^\mu \partial_\mu \phi - \mathcal{L} (\phi, \partial_\alpha \phi, x) \label{eq:legendretransformation}
	\end{align}
transfers all information from the Lagrangian density $\mathcal{L}$ to the Hamilton density $\mathcal{H}$, provided that the Hesse matrix
	\begin{align}
		\mathcal{M} & \equiv \frac{\partial^2 \mathcal{L}}{\partial(\partial_\mu \phi) \partial(\partial_\nu \phi)} \label{eq:hessematrix}
	\end{align}
is not singular.\footnote{Note that in general the conventional Dirac Lagrangian does not fulfil this requirement. Nonetheless there is a possible way out of this problem via a slightly modification of the Dirac Lagrangian, see \cite{gasiorowicz66,struckmeier08}.}
\\
In complete analogy to point mechanics it is possible to derive the corresponding canonical equations of motion (\ref{eq:canonicalone}) and (\ref{eq:canonicaltwo}) by inserting Eq.\ (\ref{eq:legendretransformation}) in the action integral (\ref{eq:action}) and vary the action, see App.\ \ref{app:canonicalequations}. This can also be seen by realizing that the Lagrange density $\mathcal{L}$ does not depend on the canonical momentum $\pi^\mu$, whereas the Hamilton density $\mathcal{H}$ does not depend on the derivatives of the field $\partial_\mu \phi$: Then Eqs.\ (\ref{eq:eulerlagrangeequation}), (\ref{eq:canonicallyconjugatefield}) and (\ref{eq:legendretransformation}) yield the canonical equations
	\begin{align}
		\frac{\partial \mathcal{H}}{\partial \phi} & = - \frac{\partial \mathcal{L}}{\partial \phi} = - \partial_\mu \frac{\partial \mathcal{L}}{\partial (\partial_\mu \phi)} = - \partial_\mu \pi^\mu \,, \label{eq:canonicalone}
		\\
		\frac{\partial \mathcal{H}}{\partial \pi^\mu} & = \partial_\mu \phi \,, \label{eq:canonicaltwo} \vphantom{\partial_\mu \frac{\partial \mathcal{L}}{\partial (\partial_\mu \phi)}}
		\\
		\left. \partial_\mu \mathcal{H}\right|_{expl} & = - \left. \partial_\mu \mathcal{L}\right|_{expl} \,. \vphantom{\partial_\mu \frac{\partial \mathcal{L}}{\partial (\partial_\mu \phi)}} \label{eq:canonicalthree}
	\end{align}
This set of two first order differential equations is equivalent to the Euler-Lagrange equation and manifestly covariant.
\\
In Sec.\ \ref{sec:lagrangianfieldtheory} and the corresponding appendices \ref{app:surfaceterms} and \ref{app:pointtransformations} we showed that Lagrange densities are not unique and a full class of them renders the same physics for the same system. It is obvious that these transformations given by Eqs.\ (\ref{eq:surfaceterms}) and (\ref{eq:pointtransformationsforfields}) should also leave the canonical equations (\ref{eq:canonicalone}) and (\ref{eq:canonicaltwo}) invariant, since Lagrangian and Hamilton densities are connected via Legendre transformations:
\\
\\
1. In App.\ \ref{app:hamiltonsurface terms} it is shown that additional surface terms in the Lagrange densities (\ref{eq:surfaceterms}) coincide with additional partial derivatives in the Hamilton densities, which do both leave the dynamics invariant,
	\begin{align}
		\mathcal{H}^\prime (\phi,\, \pi^\alpha,\, x) & = \mathcal{H}(\phi,\, \pi^\alpha,\, x) + \left. \partial_\mu \mathcal{F}^\mu (\phi,\, x)\right|_{expl} \,. \label{eq:hamiltonsurfaceterms}
	\end{align}
2. Point transformations (\ref{eq:pointtransformationsforfields}) also leave the canonical equations (\ref{eq:canonicalone}), (\ref{eq:canonicaltwo}) and therefore all physics invariant. This is derived in App.\ \ref{app:pointtransformationsham}. A point transformation in the Lagrangian picture reads
	\begin{align}
		\tilde{\mathcal{H}} (\Phi,\, \Pi^\beta,\, x) & = \mathcal{H} \left[ \phi(\Phi,\, x),\, \pi^\alpha(\Phi,\, \Pi^\beta,\, x),\, x \right] + \Pi^\mu \left. \partial_\mu \Phi \right|_{expl}
	\end{align}
in the Hamilton formulation of field theory.
\\
\\
It was already mentioned at the beginning of this chapter and at the end of Chap. \ref{sec:lagrangianfieldtheory} that there is even a higher class of transformations that leaves the dynamics of a given system invariant, whereas point transformations and surface terms are only special cases of them. These are so called canonical transformations, which will be part of the next chapter \ref{sec:canonicaltransformations}.

\subsection{Covariant canonical transformations}
\label{sec:canonicaltransformations}
In this section we derive covariant canonical transformations for a manifestly covariant Hamilton field theory, compare \cite{good54,musi78,struckmeier08}. We start with the most general transformation from Sec.\ \ref{sec:lagrangianfieldtheory} that leaves the Euler-Lagrange equation (\ref{eq:eulerlagrangeequation}) invariant, but do not restrict us any more to point transformations, which means that the differential map is formally extended to
	\begin{align}
		\Phi & = \Phi(\phi,\, \pi^\alpha,\, x) \,,
		\\
		\Pi^\mu & = \Pi^\mu(\phi,\, \pi^\alpha,\, x)
	\end{align}
and
	\begin{align}
		\phi & = \phi(\Phi,\, \Pi^\beta,\, x) \,,
				\\
		\pi^\mu & = \pi^\mu(\Phi,\, \Pi^\beta,\, x) \,.
	\end{align}
Equation (\ref{eq:lagrangians}) can be written as
	\begin{align}
		\mathcal{L}(\phi,\, \partial_\alpha \phi,\, x) & = \tilde{\mathcal{L}}(\Phi,\, \partial_\beta \Phi,\, x) + \partial_\mu f_1^\mu (\phi,\, \Phi, x) \,.
	\end{align}
Using the Legendre transformation (\ref{eq:legendretransformation}) on both sides we arrive at
	\begin{align}
		\pi^\mu \partial_\mu \phi - \mathcal{H}(\phi,\, \pi^\alpha,\, x) & = \Pi^\mu \partial_\mu \Phi - \tilde{\mathcal{H}}(\Phi,\, \Pi^\beta,\, x) + \partial_\mu f_1^\mu (\phi,\, \Phi, x) \,,
	\end{align}
which is the definition of a canonical transformation.\footnote{Note that including a scaling factor $\lambda$ would not leave phase space volume invariant and is therefore excluded in our definition of canonical transformations. In the literature transformations that include a scaling $\lambda$ are called ``extended canonical transformations''.}
\\
\\
To derive the explicit transformation rules we have to rewrite the total divergence of $f_1^\mu$,
	\begin{align}
		\pi^\mu \partial_\mu \phi - \mathcal{H}(\phi,\, \pi^\alpha,\, x) & = \Pi^\mu \partial_\mu \Phi - \tilde{\mathcal{H}}(\Phi,\, \Pi^\beta,\, x) + \frac{\partial f_1^\mu}{\partial \phi} \partial_\mu \phi + \frac{\partial f_1^\mu}{\partial \Phi} \partial_\mu \Phi + \left. \partial_\mu f_1^\mu \right|_{expl} \label{eq:kanonischetransformation}
	\end{align}
and compare coefficients to find the following rules,
	\begin{align}
		\pi^\mu & = \frac{\partial f_1^\mu}{\partial \phi} \vphantom{\frac{\partial f_1^\mu}{\partial \Phi}} \,, \label{eq:f1one}
		\\
		\Pi^\mu & = - \frac{\partial f_1^\mu}{\partial \Phi} \,, \label{eq:f1two}
		\\
		\tilde{\mathcal{H}} & = \mathcal{H} + \left. \partial_\mu f_1^\mu \right|_{expl} \vphantom{\frac{\partial f_1^\mu}{\partial \Phi}} \,. \label{eq:f1three}
	\end{align}
Note the important symmetry relation which follows from Eq.\ (\ref{eq:f1one}) and (\ref{eq:f1two}),
	\begin{align}
		\frac{\partial \pi^\mu}{\partial \Phi} & = \frac{\partial^2 f_1^\mu}{\partial \phi\, \partial \Phi} = - \frac{\partial \Pi^\mu}{\partial \phi} \,. \label{eq:f1symm}
	\end{align}
Symmetry relations have to be fulfilled by canonical transformations and therefore provide a harsh restriction on their generality, as we will see at the end of this chapter.
\\
\\
Deriving the transformation rules from Eq.\ (\ref{eq:kanonischetransformation}), we assumed that $f_1^\mu$ does only depend on $\phi,\, \Phi$ and space-time $x$. In fact there is no restriction that prevents us from having a more general surface term $f^\mu$ with a dependency on $\pi^\mu$ and $\Pi^\mu$ as well. The only restriction we have, is, that in total $f^\mu$ has always to depend on at least one old and one new variable out of $(\phi,\, \Phi,\, \pi^\mu,\, \Pi^\mu)$, since otherwise we would not have a transformation at all. In consequence there exist four different surface terms $f_1^\mu(\phi,\, \Phi,\, x),\, f_2^\mu(\phi,\, \Pi^\beta,\, x),\, f_3^\mu(\pi^\alpha,\, \Phi,\, x)$ and $f_4^\mu(\pi^\alpha,\, \Pi^\beta, x)$. These surface terms are called generating functions, due to the fact that they generate canonical transformations. We will show that they are connected via Legendre transformations.
\\
\\
In the following we will derive the generating function $f_2^\mu$: First we rewrite Eq.\ (\ref{eq:kanonischetransformation}),
	\begin{align}
		\pi^\mu \partial_\mu \phi - \mathcal{H}(\phi,\, \pi^\alpha,\, x) & = - \Phi\, \partial_\mu \Pi^\mu + \partial_\mu \left( \Pi^\mu \Phi \right) - \tilde{\mathcal{H}}(\Phi,\, \Pi^\beta,\, x) + \partial_\mu f_1^\mu (\phi,\, \Phi, x) \label{eq:f1trans}
		\\
		& = - \Phi\, \partial_\mu \Pi^\mu - \tilde{\mathcal{H}}(\Phi,\, \Pi^\beta,\, x) + \partial_\mu \left[ f_1^\mu (\phi,\, \Phi, x) + \Pi^\mu \Phi \right] \nonumber
		\\
		& = - \Phi\, \delta^\mu_\nu \partial_\mu \Pi^\nu - \tilde{\mathcal{H}}(\Phi,\, \Pi^\beta,\, x) + \partial_\mu f_2^\mu \nonumber
	\end{align}
and find that $f_1^\mu$ and $f_2^\mu$ are related via a Legendre transformation with respect to $\Phi$ and $\Pi^\mu$,
	\begin{align}
		f_2^\mu (\phi,\, \Pi^\beta,\, x) & = \Pi^\mu \Phi + f_1^\mu (\phi,\, \Phi,\, x) \,.
	\end{align}
Next we rewrite the total divergence of $f_2^\mu$ in the last line of (\ref{eq:f1trans}),
	\begin{align}
		\pi^\mu \partial_\mu \phi - \mathcal{H}(\phi,\, \pi^\alpha,\, x) & = - \Phi\, \delta^\mu_\nu \partial_\mu \Pi^\nu - \tilde{\mathcal{H}}(\Phi,\, \Pi^\beta,\, x) + \frac{\partial f_2^\mu}{\partial \phi} \partial_\mu \phi + \frac{\partial f_2^\mu}{\partial \Pi^\nu} \partial_\mu \Pi^\nu + \left. \partial_\mu f_2^\mu \right|_{expl}.
	\end{align}
Comparing coefficients results in the transformation laws
	\begin{align}
		\pi^\mu & = \frac{\partial f_2^\mu}{\partial \phi} \vphantom{\frac{\partial f_2^\mu}{\partial \Pi^\nu}} \,, \label{eq:f2one}
		\\
		\delta^\mu_\nu \Phi & = \frac{\partial f_2^\mu}{\partial \Pi^\nu} \,, \label{eq:f2two}
		\\
		\tilde{\mathcal{H}} & = \mathcal{H} + \left. \partial_\mu f_2^\mu \right|_{expl} \vphantom{\frac{\partial f_2^\mu}{\partial \Pi^\nu}} \label{eq:f2three}
	\end{align}
and the symmetry relation,
	\begin{align}
		\frac{\partial \pi^\mu}{\partial \Pi^\nu} & = \frac{\partial^2 f_2^\mu}{\partial \phi\, \partial \Pi^\nu} = \delta^\mu_\nu \frac{\partial \Phi}{\partial \phi} \,, \label{eq:f2symm}
	\end{align}
which follows from Eqs.\ (\ref{eq:f2one}) and (\ref{eq:f2two}).
\\
\\
In order to continue with $f_3^\mu$ we can start again with Eq.\ (\ref{eq:kanonischetransformation}) and rewrite the l.h.s.,
	\begin{align}
		- \phi\,  \partial_\mu \pi^\mu - \mathcal{H}(\phi,\, \pi^\alpha,\, x) & = \Pi^\mu \partial_\mu \Phi - \tilde{\mathcal{H}}(\Phi,\, \Pi^\beta,\, x) + \partial_\mu \left[ f_1^\mu (\phi,\, \Phi, x) - \pi^\mu \phi \right] \label{eq:kanonischetrans3}
		\\
		& = \Pi^\mu \partial_\mu \Phi - \tilde{\mathcal{H}}(\Phi,\, \Pi^\beta,\, x) + \partial_\mu f_3^\mu \,, \nonumber
	\end{align}
to find that $f_3^\mu$ is also a Legendre transformation of $f_1^\mu$,
	\begin{align}
		f_3^\mu (\pi^\alpha,\, \Phi,\, x) & = - \pi^\mu \phi + f_1^\mu (\phi,\, \Phi,\, x) \,.
	\end{align}
Using this result for the total derivative of $f_3^\mu$ we get
	\begin{align}
		- \phi\, \delta^\mu_\nu \partial_\mu \pi^\nu - \mathcal{H}(\phi,\, \pi^\alpha,\, x) & = \Pi^\mu \partial_\mu \Phi - \tilde{\mathcal{H}}(\Phi,\, \Pi^\beta,\, x) + \frac{\partial f_3^\mu}{\partial \pi^\nu} \partial_\mu \pi^\nu + \frac{\partial f_3^\mu}{\partial \Phi} \partial_\mu \Phi + \left. \partial_\mu f_3^\mu \right|_{expl}
	\end{align}
and extract the transformation laws
	\begin{align}
		\delta^\mu_\nu \phi & = - \frac{\partial f_3^\mu}{\partial \pi^\nu} \,, \vphantom{\frac{\partial f_3^\mu}{\partial \Phi}} \label{eq:f3one}
		\\
		\Pi^\mu & = - \frac{\partial f_3^\mu}{\partial \Phi} \,, \vphantom{\frac{\partial f_3^\mu}{\partial \Phi}} \label{eq:f3two}
		\\
		\tilde{\mathcal{H}} & = \mathcal{H} + \left. \partial_\mu f_3^\mu \right|_{expl} \,, \vphantom{\frac{\partial f_3^\mu}{\partial \Phi}} \label{eq:f3three}
	\end{align}
as well as the symmetry relation,
	\begin{align}
		\delta^\mu_\nu \frac{\partial \phi}{\partial \Phi} & = - \frac{\partial^2 f_3^\mu}{\partial \pi^\nu\, \partial \Phi} = \frac{\partial \Pi^\mu}{\partial \pi^\nu} \,, \label{eq:f3symm}
	\end{align}
compare Eqs.\ (\ref{eq:f3one}) and (\ref{eq:f3two}).
\\
\\
To find the transformation relations for generating functions $f_4^\mu$ we start with Eq.\ (\ref{eq:kanonischetrans3}) and rewrite the r.h.s.,
	\begin{align}
		- \phi\, \delta^\mu_\nu  \partial_\mu \pi^\nu - \mathcal{H}(\phi,\, \pi^\alpha,\, x) & = - \Phi\, \partial_\mu \Pi^\mu - \tilde{\mathcal{H}}(\Phi,\, \Pi^\beta,\, x) + \partial_\mu \left( f_3^\mu + \Pi^\mu \Phi \right) \vphantom{\frac{\partial f_4^\mu}{\partial \Pi^\nu}}
		\\
		& = - \Phi\, \partial_\mu \Pi^\mu - \tilde{\mathcal{H}}(\Phi,\, \Pi^\beta,\, x) + \partial_\mu f_4^\mu \vphantom{\frac{\partial f_4^\mu}{\partial \Pi^\nu}} \nonumber
		\\
		& = - \Phi\, \delta^\mu_\nu \partial_\mu \Pi^\nu - \tilde{\mathcal{H}}(\Phi,\, \Pi^\beta,\, x) + \frac{\partial f_4^\mu}{\partial \pi^\nu} \partial_\mu \pi^\nu + \frac{\partial f_4^\mu}{\partial \Pi^\nu} \partial_\mu \Pi^\nu + \left. \partial_\mu f_4^\mu \right|_{expl} \,. \nonumber
	\end{align}
Therefore $f_4^\mu$ can be identified as the Legendre transformation of $f_3^\mu$
	\begin{align}
		f_4^\mu (\pi^\alpha,\, \Pi^\beta,\, x) & = \Pi^\mu \Phi + f_3^\mu (\pi^\alpha,\, \Phi,\, x)
	\end{align}
and the transformation laws are found again via comparing the coefficients
	\begin{align}
		\delta^\mu_\nu \phi & = - \frac{\partial f_4^\mu}{\partial \pi^\nu} \vphantom{\frac{\partial f_4^\mu}{\partial \Pi^\nu}} \,, \label{eq:f4one}
		\\
		\delta^\mu_\nu \Phi & = \frac{\partial f_4^\mu}{\partial \Pi^\nu} \,, \label{eq:f4two}
		\\
		\tilde{\mathcal{H}} & = \mathcal{H} + \left. \partial_\mu f_4^\mu \right|_{expl} \vphantom{\frac{\partial f_4^\mu}{\partial \Pi^\nu}} \,. \label{eq:f4three}
	\end{align}
The last symmetry relation, that follows form Eqs.\ (\ref{eq:f4one}) and (\ref{eq:f4two}), reads
	\begin{align}
		\delta^\mu_\alpha \frac{\partial \phi}{\partial \Pi^\beta} & = - \frac{\partial^2 f_4^\mu}{\partial \pi^\alpha\, \partial \Pi^\beta} = - \delta^\mu_\beta \frac{\partial \Phi}{\partial \pi^\alpha} \,. \label{eq:f4symm}
	\end{align}
This symmetry relation has a particular property: There are three possibilities of contracting the open indices, whereas two of them lead to hard constraints for canonical transformations in general. The contraction of $\alpha$ and $\beta$ does not yield anything special, whereas contracting $\mu$ and $\beta$ leads to
	\begin{align}
		\frac{\partial \phi}{\partial \Pi^\alpha} & = - 4\, \frac{\partial \Phi}{\partial \pi^\alpha} \,,
	\end{align}
and the contraction of $\mu$ and $\alpha$ results in
	\begin{align}
		4\, \frac{\partial \phi}{\partial \Pi^\alpha} & = - \frac{\partial \Phi}{\partial \pi^\alpha} \,.
	\end{align}
We immediately see that both equations are only fulfilled simultaneously if and only if
	\begin{align}
		\frac{\partial \phi}{\partial \Pi^\alpha} & \overset{!}{=} 0 \,,
		\\
		\frac{\partial \Phi}{\partial \pi^\alpha} & \overset{!}{=} 0 \,.
	\end{align}
This is a very strict criterion for canonical transformations in general and in the realm of field theory. It states that transformed fields $\Phi$ must not depend on original momentum fields $\pi^\mu$ and original fields $\phi$ must not depend on transformed momentum fields $\Pi^\mu$,
	\begin{align}
	\phi & = \phi(\Phi,\,x) \,,
	\\
	\Phi & = \Phi(\phi,\,x) \,.
	\end{align}
This condition is consistent with the other symmetry relations (\ref{eq:f1symm}), (\ref{eq:f2symm}) and (\ref{eq:f3symm}). The inverse is not necessarily fulfilled. Original momentum fields $\pi^\mu$ can still depend on transformed fields $\Phi$ and transformed momentum fields $\Pi^\mu$ can still depend on original fields $\phi$. Although we were not able to find a suitable example for such a ``crossover'' transformation in field theory, the dependencies for $\pi^\mu$ and $\Pi^\mu$ are
	\begin{align}
	\pi^\mu & = \pi^\mu (\Phi,\,\Pi^\beta,\,x) \,,
	\\
	\Pi^\mu & = \Pi^\mu (\phi,\,\pi^\beta,\,x) \,.
	\end{align} 
 We conclude that canonical transformations in field theory are definitely restricted to point transformations for the fields $\phi$ and $\Phi$. Almost similar conditions and restrictions were already found by \cite{good54}.

\section{Scalar electrodynamics}
\label{sec:scalarelectrodynamics}
In this chapter we present a novel formulation of gauge theory in a covariant canonical transformation formalism (for an initial version compare \cite{struckmeier08,StrRei12}). We will introduce this approach using the example of $U(1)$-gauge theory.
\subsection{Local phase transformations as canonical transformations}
\label{sec:localphasetransformations}
In this section we construct classical electrodynamics as classical gauge field theory using the covariant Hamilton formulation and the theory of canonical transformations.
\\
We restrict ourself to complex scalar matter to simplify the discussion and focus on the gauging procedure. The generalisation to arbitrary massive fields, especially Dirac spinors is straightforward and will be briefly discussed in the outlook, see Sec.\ \ref{sec:outlook}.
\\
Therefore we start with a Hamilton density $\mathcal{H}$, which only depends on a complex scalar field $\phi,\, \bar{\phi}$, its canonically conjugate field $\bar{\pi}^\mu,\, \pi^\mu$ and the space-time position $x$,
	\begin{align}
		\mathcal{H} & \equiv \mathcal{H}(\phi,\, \bar{\phi},\, \bar{\pi}^\alpha,\, \pi^\alpha,\, x) \,. \label{eq:hamiltonian}
	\end{align}
One restriction is, that the dynamical part of the Hamilton density has to be only quadratic in $\pi^\mu$ and $\bar{\pi}^\mu$ and the Hamilton density has to be hermitian, which means
	\begin{align}
		\mathcal{H} & = \bar{\pi}_\mu \pi^\mu + V(\phi,\, \bar{\phi},\, x) \,, \label{eq:hamiltonstart}
	\end{align}
where $V$ denotes a potential term, including mass, self interaction terms and direct space-time dependencies.
\\
\\
We require this matter Hamiltonian to be invariant under the global gauge transformation
	\begin{align}
		\phi \longrightarrow \Phi & = e^{-i\Lambda}\, \phi \,,
		\\
		\bar{\phi} \longrightarrow \bar{\Phi} & = \bar{\phi}\, e^{i\Lambda} \,,
	\end{align}
with a real global phase factor $\Lambda$. (The transformation properties of $\pi^\mu$ and $\bar{\pi}^\mu$ are not clear at this point, but also the dynamic part $\bar{\pi}_\mu \pi^\mu$ of Eq.\ (\ref{eq:hamiltonstart}) has to be invariant under global phase transformations.)
\\
\\
The request that the invariance of $\mathcal{H}$ is retained even under so called local gauge transformations
	\begin{align}
		\phi \longrightarrow \Phi & = e^{-i\Lambda(x)}\, \phi \,, \label{eq:transformationphi}
		\\
		\bar{\phi} \longrightarrow \bar{\Phi} & = \bar{\phi}\, e^{i\Lambda(x)} \,, \label{eq:transformationphibar}
	\end{align}
with any real phase function $\Lambda (x)$ is a far reaching $U(1)$-symmetry requirement driving the introduction of gauge fields. (For notational simplicity we will drop the $x$-dependency of $\Lambda$ in all following considerations.) 
\\
In the following we show how a consistent gauge field theory emerges naturally from that local symmetry in the framework of the canonical transformaion theory. We will use the generating function $f^\mu_2 (\phi,\, \Pi^\mu,\, x)$ encountered already in Sec.\ \ref{sec:canonicaltransformations} which for complex scalar fields has the form\footnote{Note that we could have also chosen $f_3^\mu$ instead of $f_2^\mu$, since both are eligible to generate point transformations.}
	\begin{align}
		f^\mu_2 & \equiv f^\mu_2 (\phi,\, \bar{\phi},\, \Pi^\beta,\, \bar{\Pi}^\beta,\, x)\,,
	\end{align}
where $\Pi^\mu$ and $\bar{\Pi}^\mu$ are defined as the canonical conjugate fields of $\bar{\Phi}$ and $\Phi$, compare Sec.\ \ref{sec:canonicaltransformations}.
\\
A proper generating function $f_2^\mu$ can be found if we claim that Eq.\ (\ref{eq:f2two}) has to reproduce the local phase transformations (\ref{eq:transformationphi}) and (\ref{eq:transformationphibar}) of the fields $\phi$ and $\bar{\phi}$,
	\begin{align}
		\delta^\mu_\nu \Phi & = \frac{\partial f_2^\mu}{\partial \bar{\Pi}^\nu} \overset{!}{=} \delta^\mu_\nu\,  e^{-i\Lambda}\, \phi \,, \label{eq:transphi}
		\\
		\delta^\mu_\nu \bar{\Phi} & = \frac{\partial f^\mu_2}{\partial \Pi^\nu} \overset{!}{=} \delta^\mu_\nu\, \bar{\phi}\, e^{i\Lambda} \,. \label{eq:transphibar}
	\end{align}
Partial functional-integrating Eqs.\ (\ref{eq:transphi}) and (\ref{eq:transphibar}) with respect to $\bar{\Pi}^\nu$ and $\Pi^\nu$, we find
	\begin{align}
		f^\mu_2 & = \bar{\Pi}^\mu e^{-i\Lambda}\, \phi + \bar{\phi}\, e^{i\Lambda} \Pi^\mu + c^\mu(\phi,\, \bar{\phi},\, x) \,, \label{eq:f2firstapproach}
	\end{align}
where $c^\mu(\phi,\, \bar{\phi},\, x)$ is an arbitrary integration constant/integration function. This generating function does now reproduce local phase transformations of massive scalar fields.
\\
\\
According to Eq.\ (\ref{eq:f2one}) we can also derive the transformation rules for the canonically conjugate fields, which read
	\begin{align}
		\pi^\mu & = \frac{\partial f^\mu_2}{\partial \bar{\phi}} = e^{i\Lambda} \Pi^\mu + \frac{\partial c^\mu}{\partial \bar{\phi}} \,, \label{eq:transmomentum}
		\\
		\bar{\pi}^\mu & = \frac{\partial f^\mu_2}{\partial \phi} = \bar{\Pi}^\mu e^{-i\Lambda} + \frac{\partial c^\mu}{\partial \phi} \,. \label{eq:transmomentumbar}
	\end{align}
Let us stop here for a moment and have a closer look at the generating function (\ref{eq:f2firstapproach}):
\\
\\
In order to check whether these transformations are canonical, one has to verify that all symmetry relations (\ref{eq:f1symm}), (\ref{eq:f2symm}), (\ref{eq:f3symm}) and (\ref{eq:f4symm}) are fulfilled and, in consequence, the phase-space volume remains constant. All symmetry relations are calculated in App.\ \ref{app:proofcanonicaltrans}. We find that the symmetry relations restrict our choice for the integration constant $c^\mu$ strongly. In fact only functions that are linear in $\bar{\phi}$ and $\phi$ with arbitrary $x$-dependency are still allowed,
	\begin{align}
		c^\mu (\phi,\, \bar{\phi},\, x) & = u^\mu (x)\, \phi + v^\mu (x)\, \bar{\phi} + w^\mu (x) \,, \label{eq:integrationkonst12}
	\end{align}
where $u^\mu$, $v^\mu$ and $w^\mu$ are as yet arbitrary $x$-dependent functions (we also drop the argument $x$ in the following for $u^\mu$, $v^\mu$ and $w^\mu$). We will explain the importance of $w^\mu$ later, but already mention here that it can be chosen arbitrary. In contrast to $w^\mu$, the functions $u^\mu$ and $v^\mu$ have direct influence on the transformation rules for the canonically conjugate fields (\ref{eq:transmomentum}) and (\ref{eq:transmomentumbar})
	\begin{align}
		\pi^\mu & = \frac{\partial f^\mu_2}{\partial \bar{\phi}} = e^{i\Lambda} \Pi^\mu + v^\mu \,, \label{eq:pitrafo}
		\\
		\bar{\pi}^\mu & = \frac{\partial f^\mu_2}{\partial \phi} = \bar{\Pi}^\mu e^{-i\Lambda} + u^\mu \,. \label{eq:barpitrafo}
	\end{align}
However, in the beginning of this section we demanded that the Hamilton density (\ref{eq:hamiltonstart}) has to be invariant under global transformations. Let us therefore insert the transformation rules for $\pi^\mu$ (\ref{eq:pitrafo}) and $\bar{\pi}^\mu$ (\ref{eq:barpitrafo}) as well as $\phi$ (\ref{eq:transphi}) and $\bar{\phi}$ (\ref{eq:transphibar}) in (\ref{eq:hamiltonstart}) for $\Lambda = const.$ and assume that $V(\phi,\, \bar{\phi},\, x)$ is invariant under global phase transformations on its own (which completely makes sense for a mass or self-interaction term),
	\begin{align}
		\mathcal{H} & = \bar{\pi}_\mu \pi^\mu + V(\phi,\, \bar{\phi},\, x)
		\\
		& = \left( \bar{\Pi}_\mu e^{-i\Lambda} + u_\mu \right) \left( e^{i\Lambda} \Pi^\mu + v^\mu \right) + V(\Phi,\, \bar{\Phi},\, x) \nonumber
		\\
		& = \bar{\Pi}_\mu \Pi^\mu + \bar{\Pi}_\mu v^\mu e^{-i\Lambda} + u_\mu \Pi^\mu  e^{i\Lambda} + u_\mu v^\mu + V(\Phi,\, \bar{\Phi},\, x) \nonumber 
		\\
		& = \mathcal{H}^\prime + \bar{\Pi}_\mu v^\mu e^{-i\Lambda} + u_\mu \Pi^\mu  e^{i\Lambda} + u_\mu v^\mu \,. \nonumber
	\end{align}
Note: In fact there is no choice (except: both are zero) for $u^\mu$ and $v^\nu$ which does only depend on $x$ that makes the second and third term vanish or cancel each other. The third term does not matter, since it only depends on $x$ and leaves the canonical equations (\ref{eq:canonicalone}) and (\ref{eq:canonicaltwo}) invariant.\\
We conclude that we have to choose $u^\mu = v^\nu = 0$ to conserve global invariance under phase transformations of the Hamilton density. Therefore $f_2^\mu$ reads
	\begin{align}
		f^\mu_2 & = \bar{\Pi}^\mu e^{-i\Lambda}\, \phi + \bar{\phi}\, e^{i\Lambda} \Pi^\mu + w^\mu \,. \label{eq:f2}
	\end{align}
Consequently Eqs.\ (\ref{eq:transmomentum}) and (\ref{eq:transmomentumbar}) simplify to
	\begin{align}
		\pi^\mu & = \frac{\partial f^\mu_2}{\partial \bar{\phi}} = e^{i\Lambda} \Pi^\mu \,, \label{eq:transmomentumneu}
		\\
		\bar{\pi}^\mu & = \frac{\partial f^\mu_2}{\partial \phi} = \bar{\Pi}^\mu e^{-i\Lambda} \label{eq:transmomentumbarneu}
	\end{align}
and we are ready to continue our main discussion.
\\
\\
In the next step we use Eq.\ (\ref{eq:f2three}) to derive the transformation rule for the Hamilton density
	\begin{align}
		\mathcal{H}^\prime (\Phi,\, \bar{\Phi},\, \Pi^\beta,\, \bar{\Pi}^\beta,\, x) - \mathcal{H} (\phi,\, \bar{\phi},\, \pi^\alpha,\, \bar{\pi}^\alpha,\, x) & = \left. \partial_\mu f^\mu_2 \right|_{expl} \nonumber
		\\
		& = -i (\partial_\mu \Lambda) \left( \bar{\Pi}^\mu e^{-i\Lambda}\, \phi - \bar{\phi}\, e^{i\Lambda} \Pi^\mu \right) + \partial_\mu w^\mu  \,.
	\end{align}
Inserting the transformation rules (\ref{eq:transmomentum}) and (\ref{eq:transmomentumbar}) this reduces to
	\begin{align}
		\mathcal{H}^\prime - \mathcal{H} & = -i (\partial_\mu \Lambda) \left( \bar{\pi}^\mu\, \phi - \bar{\phi}\, \pi^\mu \right) + \partial_\mu w^\mu \,. \label{eq:transhamiltonians}
	\end{align}
In the beginning of this section we pointed out that the Lagrangian or respective Hamilton densities of a fundamental theory should be form-invariant under local gauge transformations. In fact, we find that the generating function (\ref{eq:f2}) leads to Eq.\ (\ref{eq:transhamiltonians}) which means that the Hamilton density (\ref{eq:hamiltonian}) is no longer form-invariant. In consequence the Hamilton density (\ref{eq:hamiltonian}) has to be modified in a way to recover form-invariance. At this step the actual gauging procedure begins: To compensate the additional $\Lambda$-term in Eq.\ (\ref{eq:transhamiltonians}) we introduce an additional vector field $a_\mu$. Then we define an amended Hamilton density $\mathcal{H}_a (\phi,\, \bar{\phi},\, \bar{\pi}^\alpha,\, \pi^\alpha,\, a_\alpha,\, p^{\alpha\gamma},\, x)$, where $p^{\mu\nu}$ is said to be the canonically conjugate field of the gauge  field $a_\mu$. Notice that the introduction of $p^{\mu\nu}$ is at this point only a formal introduction to make $a_\mu$ a dynamical field in phase space, but does not mean that we know the dependency of $\mathcal{H}_a$ on $p^{\mu\nu}$ yet.\\
Let us for the moment assume that $\mathcal{H}_a$ depends on the original Hamilton density and an extra term which has the same shape as the additional term in Eq.\ (\ref{eq:transhamiltonians}),
	\begin{align}
		\mathcal{H}_a & \equiv \mathcal{H} + i q \left( \bar{\pi}^\mu\, \phi - \bar{\phi}\, \pi^\mu \right) a_\mu \,. \label{eq:modhamiltonian}
	\end{align}
This choice is a minimal one which is referred to ``minimal coupling'' and therefore only sufficient, but not necessary. It will turn out that the term in the parenthesis represents the electromagnetic current density and thus acts as the source of the electromagnetic field.
\\
Notice that we have introduced the coupling constant $q$, which will be identified as the electric charge. In the next step we require this Hamilton density to remain invariant under the local phase transformations, except for the total divergence $\partial_\mu w^\mu$, that does not contribute to the equations of motion. This means,
	\begin{align}
		\mathcal{H}^\prime_a & = \mathcal{H}^\prime + i q \left( \bar{\Pi}^\mu\, \Phi - \bar{\Phi}\, \Pi^\mu \right) A_\mu - \partial_\mu w^\mu \label{eq:modhamiltonianbar}
	\end{align}
and
	\begin{align}
		0 & = \mathcal{H}^\prime_a - \mathcal{H}_a \,.
	\end{align}
Inserting the Hamilton densities (\ref{eq:modhamiltonian}) and (\ref{eq:modhamiltonianbar}) in this equation we get
	\begin{align}
		0 & = \mathcal{H}^\prime - \mathcal{H} + i q \left( \bar{\Pi}^\mu\, \Phi - \bar{\Phi}\, \Pi^\mu \right) A_\mu - i q \left( \bar{\pi}^\mu\, \phi - \bar{\phi}\, \pi^\mu \right) a_\mu - \partial_\mu w^\mu \nonumber
		\\
		& = -i (\partial_\mu \Lambda) \left( \bar{\pi}^\mu\, \phi - \bar{\phi}\, \pi^\mu \right) + i q \left( \bar{\Pi}^\mu\, \Phi - \bar{\Phi}\, \Pi^\mu \right) A_\mu - i q \left( \bar{\pi}^\mu\, \phi - \bar{\phi}\, \pi^\mu \right) a_\mu \,,
	\end{align}
where we used Eq.\ (\ref{eq:transhamiltonians}) in the last line. With the transformation rules (\ref{eq:transformationphi}), (\ref{eq:transformationphibar}), (\ref{eq:transmomentumneu}) and (\ref{eq:transmomentumbarneu}), this reduces to
	\begin{align}
	0 & = \left( \bar{\pi}^\mu\, \phi - \bar{\phi}\, \pi^\mu \right) \left[ A_\mu  - a_\mu - \frac{1}{q} (\partial_\mu \Lambda)\right]
	\end{align}
and therefore leads to the transformation rule for the gauge field,
	\begin{align}
		A_\mu  = a_\mu + \frac{1}{q} (\partial_\mu \Lambda) \,. \label{eq:transgaugefield}
	\end{align}
We immediately recognize that this is exactly the well known transformation behaviour of the gauge bosons in $U(1)$-gauge theory.

\subsection{Gauge field dynamics}
\label{sec:gaugefielddynamics}
In the last section we derived an amended Hamilton density (\ref{eq:modhamiltonian})
	\begin{align}
		\mathcal{H}_a (\phi,\, \bar{\phi},\, \bar{\pi}^\alpha,\, \pi^\alpha,\, a_\alpha,\, p^{\alpha\gamma},\, x) & \equiv \mathcal{H} (\phi,\, \bar{\phi},\, \bar{\pi}^\alpha,\, \pi^\alpha,\, x) + i q \left( \bar{\pi}^\mu\, \phi - \bar{\phi}\, \pi^\mu \right) a_\mu
	\end{align}
which is form-invariant under local phase transformations. These transformations were given by a generating function $f^\mu_2$,
	\begin{align}
		f^\mu_2 & = \bar{\Pi}^\mu e^{-i\Lambda}\, \phi + \bar{\phi}\, e^{i\Lambda} \Pi^\mu + w^\mu \,.
	\end{align}
In the end we derived the transformation behaviour for the gauge fields (\ref{eq:transgaugefield}), which reads
	\begin{align}
		A_\mu  = a_\mu + \frac{1}{q} (\partial_\mu \Lambda) \,.
	\end{align}
To finish the gauging process we still have to include the transformation behaviour of the gauge fields $a_\mu$ in the generating function $f_2^\mu$. Therefore we extend the dependency of $f_2^\mu$, see Eq.\ (\ref{eq:f2}), to $f^\mu_2 (\phi,\, \bar{\phi},\, \Pi^\beta,\, \bar{\Pi}^\beta,\, a_\alpha,\, P^{\beta\delta},\, x)$ and rewrite Eq.\ (\ref{eq:f2two}) for the gauge field to claim
	\begin{align}
		\delta^\mu_\nu A_\alpha & = \frac{\partial f^\mu_2}{\partial P^{\alpha\nu}} \overset{!}{=} \delta^\mu_\nu \left[ a_\alpha + \frac{1}{q} (\partial_\alpha \Lambda) \right] \,, \label{eq:transgauge}
	\end{align}
where we defined $P^{\alpha\nu}$ as the canonically conjugate field of the transformed gauge field $A_\alpha$. Partial functional-integrating with respect to $P^{\alpha\nu}$ leads to
	\begin{align}
		f^\mu_2 & = \bar{\Pi}^\mu e^{-i\Lambda}\, \phi + \bar{\phi}\, e^{i\Lambda} \Pi^\mu + P^{\alpha\mu} \left[ a_\alpha + \frac{1}{q} (\partial_\alpha \Lambda) \right] + c^{\prime\mu} (a_\alpha, x) \,,
	\end{align}
where $c^{\prime\mu}$ includes $w^\mu$.\\
\\
In analogy to the previous discussion in Sec.\ \ref{sec:localphasetransformations} we first check whether this $f_2^\mu$ is actually a canonical transformation. Therefore we calculate the transformation law for $p^{\nu\mu}$,
	\begin{align}
		p^{\nu\mu} & = \frac{\partial f_2^\mu}{\partial a_\nu} = P^{\nu\mu} + \frac{\partial c^{\prime \mu}}{\partial a_\nu} \label{eq:transkonjgauge}
	\end{align}
and test the symmetry relations in App.\ \ref{app:symmetryforgauge}. The symmetry relations (\ref{eq:f1symm}), (\ref{eq:f2symm}), (\ref{eq:f3symm}) and (\ref{eq:f4symm}) are converted for gauge fields. We find that the integration constant $c^{\prime\mu}$ can only be a linear function in $a_\mu$ and $x$ dependent to be compatible with the symmetry relations,
	\begin{align}
		c^{\prime\mu} (a_\alpha,\, x) & = y^{\nu\mu} (x)\, a_\nu + w^\mu (x) \,,
	\end{align}
where $y^{\mu\nu}$ is an $x$-dependent function and $w^\mu$ is already known from the previous section. (We omit all $x$- and $a_\alpha$-dependencies again.) In consequence the transformation law (\ref{eq:transkonjgauge}) for the canonically conjugate field $p^{\nu\mu}$ is
	\begin{align}
		p^{\nu\mu} & = P^{\nu\mu} + y^{\nu\mu} \,, \label{eq:kleinpzugrossp}
	\end{align}
where $y^{\nu\mu}$ is arbitrary. This arbitrariness is the last gap in a continuous derivation of a gauge theory from basic principles. We are confident though that there is less freedom than we have encountered so far. (In App.\ \ref{app:integrationconstant}, we provide a motivation for this specific choice of the integration constant $y^{\nu\mu}$ by making further assumptions about the final theory.)
\\
In the following considerations we choose $y^{\mu\nu}$ and $w^\mu$ as
	\begin{align}
		y^{\nu\mu} & = \frac{1}{q} \left[ -\eta^{\nu\mu} (\Box \Lambda) + (\partial^\nu \partial^\mu \Lambda) \right] \,, \label{eq:problemgleichung}
		\\
		w^\mu & = \frac{1}{2\, q} (\partial_\nu \Lambda)\, y^{\nu\mu} + w^{\prime\mu} \,,
	\end{align}
where $w^{\prime\mu}$ is an arbitrary function of $x$. In consequence we have to show that at least the choice of $w^\mu$ does not violate generality: An arbitrary choice of $w^\mu$ does not influence dynamics at all, since $w^\mu$ does only depend on $x$ and hence does not contribute to the canonical equations! Therefore we can split $w^\mu$ up into a specific choice and a still arbitrary function of $x$, e.g. $w^{\prime\mu}$.
\\
\\
We continue with our new generating function $f_2^\mu$,
	\begin{align}
		f^\mu_2 & = \bar{\Pi}^\mu e^{-i\Lambda(x)}\, \phi + \bar{\phi}\, e^{i\Lambda(x)} \Pi^\mu \vphantom{\left[ a_\mu + \frac{1}{e} (\partial_\mu \Lambda) \right]} + P^{\alpha\mu} \left[ a_\alpha + \frac{1}{q} (\partial_\alpha \Lambda) \right]
		\\
		& \quad + \left[a_\alpha + \frac{1}{2\, q} (\partial_\alpha \Lambda) \right] \frac{1}{q} \left[ -\eta^{\alpha\mu} (\Box \Lambda) + (\partial^\alpha \partial^\mu \Lambda) \right] + w^{\prime\mu} \,, \nonumber
	\end{align}
which still leads to the correct transformation behaviour for the gauge fields $a_\mu$, since the additional terms in the second line do not depend on $P^{\mu\nu}$.
\\
In contrast to Sec.\ \ref{sec:localphasetransformations}, this integration function modifies the transformation rules for the canonically conjugate field $p^{\mu\nu}$ of the gauge field due to its $a_\alpha$-dependence,
	\begin{align}
		p^{\nu\mu} & = \frac{\partial f^\mu_2}{\partial a_\nu} = P^{\nu\mu} + \frac{1}{q} \left[ -\eta^{\nu\mu} (\Box \Lambda) + (\partial^\nu \partial^\mu \Lambda) \right] \,. \label{eq:transcancongauge}
	\end{align}
Notice that the additional term has vanishing divergence and has therefore the structure of a projection operator. The trace $(p \equiv \tensor{p}{^\mu_\mu})$ of this equation yields
	\begin{align}
		p & = P - \frac{3}{q} (\Box \Lambda) \,. \label{eq:transtrace}
	\end{align}
Combining Eqs.\ (\ref{eq:transcancongauge}) and (\ref{eq:transtrace}) we find
	\begin{align}
		 p^{\nu\mu} - \frac{1}{3} \eta^{\nu\mu}\, p - P^{\nu\mu} + \frac{1}{3} \eta^{\nu\mu} P & = \frac{1}{q} \partial^\nu \partial^\mu \Lambda \,, \label{eq:usefultool}
	\end{align}
which will be used in the following. However, the specific choice of $y^{\nu\mu}$ has only influence on the transformation behaviour of the symmetric parts in $p^{\nu\mu}$. The antisymmetric parts of the canonically conjugate field transform homogeneously without any influence from the integration constants.
\\ 
\\
The transformation rule of the Hamilton densities is modified as well, since according to the general rule (\ref{eq:f2three}),
	\begin{align}
		\mathcal{H}^\prime - \mathcal{H} & = \left. \partial_\mu f^\mu_2 \right|_{expl} \vphantom{\frac{1}{2\, e^2}} \label{eq:hamtransnew}
		\\
		& = -i (\partial_\mu \Lambda) \left( \bar{\Pi}^\mu e^{-i\Lambda}\, \phi - \bar{\phi}\, e^{i\Lambda} \Pi^\mu \right) + \frac{1}{q} P^{\mu\nu} \left( \partial_\mu \partial_\nu \Lambda \right) + \frac{1}{q} a_\alpha \left[ - (\partial^\alpha \Box \Lambda) + (\Box \partial^\alpha \Lambda) \right] \vphantom{\frac{1}{2\, e^2}} \nonumber
		\\
		& \quad + \frac{1}{2\, q^2} (\partial_\mu \partial_\alpha \Lambda) \left[ -\eta^{\alpha\mu} (\Box \Lambda) + (\partial^\alpha \partial^\mu \Lambda) \right] + \frac{1}{2\, q^2} (\partial_\alpha \Lambda) \left[ - (\Box \partial^\alpha \Lambda) + (\Box \partial^\alpha \Lambda) \right] + \partial_\mu w^{\prime\mu} \vphantom{\frac{1}{2\, e^2}} \nonumber
		\\
		& = iq (a_\mu - A_\mu) \left( \bar{\Pi}^\mu e^{-i\Lambda}\, \phi - \bar{\phi}\, e^{i\Lambda} \Pi^\mu \right) + \frac{1}{q} P^{\mu\nu} \left( \partial_\mu \partial_\nu \Lambda \right) \vphantom{\frac{1}{2\, e^2}} \nonumber
		\\
		& \quad + \frac{1}{2\, q^2} \left[ - (\Box \Lambda) (\Box \Lambda) + (\partial_\alpha \partial_\mu \Lambda) (\partial^\alpha \partial^\mu \Lambda) \right] + \partial_\mu w^{\prime\mu} \vphantom{\frac{1}{2\, e^2}} \nonumber
		\\
		& = iq \left( \bar{\pi}^\mu \phi - \bar{\phi}\, \pi^\mu \right) a_\mu - iq \left( \bar{\Pi}^\mu \Phi - \bar{\Phi}\, \Pi^\mu \right) A_\mu + \frac{1}{q} P^{\mu\nu} \left( \partial_\mu \partial_\nu \Lambda \right) \vphantom{\frac{1}{2\, e^2}} \nonumber
		\\
		& \quad + \frac{1}{2\, q^2} \left[ - (\Box \Lambda) (\Box \Lambda) + (\partial_\alpha \partial_\mu \Lambda) (\partial^\alpha \partial^\mu \Lambda) \right] + \partial_\mu w^{\prime\mu} \,. \vphantom{\frac{1}{2\, e^2}} \nonumber
	\end{align}
At this point we can use relation (\ref{eq:transtrace}) and (\ref{eq:usefultool}) to simplify the last expression,
	\begin{align}
		\mathcal{H}^\prime - \mathcal{H} & = iq \left( \bar{\pi}^\mu \phi - \bar{\phi}\, \pi^\mu \right) a_\mu - iq \left( \bar{\Pi}^\mu \Phi - \bar{\Phi}\, \Pi^\mu \right) A_\mu \vphantom{\left( \frac{1}{18} \right)^2} \nonumber
		\\
		& \quad + P^{\nu\mu} \left[ \left( p_{\nu\mu} - \frac{1}{3} \eta_{\nu\mu}\, p \right) - \left( P_{\nu\mu} - \frac{1}{3} \eta_{\nu\mu} P \right) \right] - \frac{1}{18} \left(p- P\right)^2 \vphantom{\left( \frac{1}{18} \right)^2} \nonumber
		\\
		& \quad + \frac{1}{2} \left[ \left( p^{\nu\mu} - \frac{1}{3} \eta^{\nu\mu}\, p \right) - \left( P^{\nu\mu} - \frac{1}{3} \eta^{\nu\mu} P \right) \right]^2 + \partial_\mu w^{\prime\mu} \vphantom{\left( \frac{1}{18} \right)^2} \nonumber
		\\
		& = iq \left( \bar{\pi}^\mu \phi - \bar{\phi}\, \pi^\mu \right) a_\mu - iq \left( \bar{\Pi}^\mu \Phi - \bar{\Phi}\, \Pi^\mu \right) A_\mu \vphantom{\left( \frac{1}{18} \right)^2} \nonumber
		\\
		& \quad + \frac{1}{2} p_{\nu\mu} p^{\nu\mu} - \frac{1}{6} p^2 - \frac{1}{2} P_{\nu\mu} P^{\nu\mu} + \frac{1}{6} P^2 + \partial_\mu w^{\prime\mu} \vphantom{\left( \frac{1}{18} \right)^2} \,. \nonumber
	\end{align}
After reordering all terms we find
	\begin{align}
		\mathcal{H}^\prime + iq \left( \bar{\Pi}^\mu \Phi - \bar{\Phi}\, \Pi^\mu \right) A_\mu + \frac{1}{2} P_{\beta\alpha} P^{\beta\alpha} - \frac{1}{6} P^2 & = \mathcal{H} + iq \left( \bar{\pi}^\mu \phi - \bar{\phi}\, \pi^\mu \right) a_\mu \label{eq:trolo}
		\\
		& \quad + \frac{1}{2} p_{\beta\alpha} p^{\beta\alpha} - \frac{1}{6} p^2 + \partial_\mu w^{\prime\mu} \,. \nonumber
	\end{align}
We immediately see that the transformed Hamiltonian on the l.h.s. has exactly the same form as the original Hamiltonian on the r.h.s., up to an arbitrary $x$-dependent function $\partial_\mu w^{\prime\mu}$. Since this $x$-dependent function does not contribute to the dynamics and can always be neglected. In consequence we have found a form invariant Hamilton density.
\\
In total the amended Hamiltonian $\mathcal{H}_a (\phi,\, \bar{\phi},\, \bar{\pi}^\alpha,\, \pi^\alpha,\, a_\alpha,\, p^{\alpha\gamma},\, x)$ for classical scalar electrodynamics is therefore given by
	\begin{align}
		\mathcal{H}_a (\phi,\, \bar{\phi},\, \bar{\pi}^\alpha,\, \pi^\alpha,\, a_\alpha,\, p^{\alpha\gamma},\, x) & = \mathcal{H} + iq \left( \bar{\pi}^\mu \phi - \bar{\phi}\, \pi^\mu \right) a_\mu + \frac{1}{2} p_{\beta\alpha} p^{\beta\alpha} - \frac{1}{6} p^2 \,. \label{eq:hamiltonianfinal}
	\end{align}
At this point we have to note that we cannot be sure whether further terms could be added to the Hamiltonian by hand which are gauge invariant on their own. One term which fulfils this property is the so called ``$\theta$-term''. It is gauge invariant for all symmetric choices of $y^{\nu\mu}$
	\begin{align}
		\epsilon_{\mu\nu\alpha\beta}\, p^{\mu\nu} p^{\alpha\beta} & = \epsilon_{\mu\nu\alpha\beta}\, (P^{\mu\nu} + y^{\mu\nu}) (P^{\alpha\beta} + y^{\alpha\beta})
		\\
		& = \epsilon_{\mu\nu\alpha\beta}\, P^{\mu\nu} P^{\alpha\beta} \,. \nonumber
	\end{align}
Adding such a term (or powers of it) to the final amended Hamiltonian (\ref{eq:hamiltonianfinal}) is like adding a null to Eq.\ (\ref{eq:trolo}). It cannot be part of the canonical transformations since it has no contribution to the generating function. Neglecting the massive scalar field in Eq.\ (\ref{eq:hamiltonianfinal}) it is easy to show that this specific term does not contribute to the dynamics of the gauge fields, whereas in general, or for higher orders this cannot be excluded. The only argument that can exclude these terms in a non physical argumentation is ``Occam's razor''.

\subsection{Klein-Gordon-Maxwell system - in the Hamilton picture}
\label{sec:kgmaxwellhamilton}
In this section we show that the deduced Hamilton density (\ref{eq:hamiltonianfinal}) indeed reproduces scalar electrodynamics if we choose the complex Klein-Gordon field for the matter field \cite{greiner96}. Therefore $\mathcal{H}$ is given by Eq.\ (\ref{eq:kleingordonhamiltonian}) and the amended Hamiltonian (\ref{eq:hamiltonianfinal}) reads (we drop the Index ``$a$'')
	\begin{align}
		\mathcal{H} & = \bar{\pi}_\mu \pi^\mu + m^2 \bar{\phi}\, \phi + iq \left( \bar{\pi}^\mu \phi - \bar{\phi}\, \pi^\mu \right) a_\mu + \frac{1}{2} p_{\beta\alpha} p^{\beta\alpha} - \frac{1}{6} p^2 \,. \label{eq:hamiltonfinalkg}
	\end{align}
In the first step we calculate the canonical equations (\ref{eq:canonicalone}) and (\ref{eq:canonicaltwo}) for the matter fields $\bar{\phi}$ and $\phi$,
	\begin{align}
		- \partial_\mu \bar{\pi}^\mu & = \frac{\partial \mathcal{H}}{\partial \phi} = m^2 \bar{\phi} + iq\, \bar{\pi}^\mu\, a_\mu \vphantom{\frac{\partial \mathcal{H}}{\partial \bar{\phi}}} \,,
		\\
		\partial_\mu \bar{\phi} & = \frac{\partial \mathcal{H}}{\partial \pi^\mu} = \bar{\pi}_\mu - iq\, \bar{\phi}\, a_\mu \vphantom{\frac{\partial \mathcal{H}}{\partial \bar{\phi}}} \,, \label{eq:legbarpi}
		\\
		- \partial_\mu \pi^\mu & = \frac{\partial \mathcal{H}}{\partial \bar{\phi}} = m^2 \phi -iq\, \pi^\mu\, a_\mu \vphantom{\frac{\partial \mathcal{H}}{\partial \bar{\phi}}} \,,
		\\
		\partial_\mu \phi & = \frac{\partial \mathcal{H}}{\partial \bar{\pi}^\mu} = \pi_\mu + iq\, \phi\, a_\mu \vphantom{\frac{\partial \mathcal{H}}{\partial \bar{\phi}}} \,. \label{eq:legpi}
	\end{align}
We define the gauge covariant derivative $\mathcal{D}_\mu$ as follows
	\begin{align}
		\mathcal{D}^{*}_\mu & \equiv \partial_\mu + iq\, a_\mu \,,
		\\
		\mathcal{D}_\mu & \equiv \partial_\mu - iq\, a_\mu \,,
	\end{align}
and find via combining the above equations:
	\begin{align}
		0 & = (\mathcal{D}^{*}_\mu \mathcal{D}^{* \mu} + m^2)\, \bar{\phi} \,, \label{eq:kgmaxwelleq}
		\\
		0 & = (\mathcal{D}_\mu \mathcal{D}^\mu + m^2)\, \phi \,, \label{eq:kgmaxwelleq2}
	\end{align}
which are the well known Klein-Gordon-Maxwell equations, that describe the relativistic evolution of a complex scalar field in the presence of an electromagnetic field.
\\
\\
In the next step we calculate the canonical equations (\ref{eq:canonicalone}) and (\ref{eq:canonicaltwo}) for the electromagnetic field/the gauge field:
	\begin{align}
		- \partial_\mu p^{\nu\mu} & = \frac{\partial \mathcal{H}}{\partial a_\nu} = iq\, (\bar{\pi}^\nu \phi - \bar{\phi} \pi^\nu) \label{eq:gaugecanone}
		\\
		& = iq\, \left\{ \left[ (\partial^\nu \bar{\phi}) + iq\, \bar{\phi}\, a^\nu \right] \phi - \bar{\phi} \left[ (\partial^\nu \phi) - iq\, \phi\, a^\nu \right] \right\} \vphantom{\frac{\partial \mathcal{H}}{\partial a_\nu}} \nonumber
		\\
		& = - j^\nu - 2\,q^2\, \bar{\phi}\, \phi\, a^\nu \vphantom{\frac{\partial \mathcal{H}}{\partial a_\nu}} \,, \nonumber
	\end{align}
where
	\begin{align}
		j^\nu & \equiv iq\, \left[ \bar{\phi}\, (\partial^\nu \phi) - (\partial^\nu \bar{\phi})\, \phi \right] \label{eq:currentdensity}
	\end{align}
is the current density.
	\begin{align}
		\partial_\mu a_\nu & = \frac{\partial \mathcal{H}}{\partial p^{\nu\mu}} = p_{\nu\mu} - \frac{1}{3} \eta_{\nu\mu} p \,. \label{eq:gaugecantwo}
	\end{align}
Taking the trace of this equation results in
	\begin{align}
		\partial_\mu a^\mu & = - \frac{1}{3} p \,, \label{eq:gaugecantwotrace}
	\end{align}
which can be used to calculate the derivatives of Eq.\ (\ref{eq:gaugecantwo}) with respect to $\mu$ and $\nu$,
	\begin{align}
		\partial_\nu \partial^\mu a^\nu & = \partial_\nu p^{\nu\mu} - \frac{1}{3} \partial^\mu p
		\\
		& = \partial_\nu p^{\nu\mu} + \partial^\mu \partial_\nu a^\nu \vphantom{\frac{1}{3}} \longrightarrow 0 = \partial_\nu p^{\nu\mu} \,, \nonumber
		\\
		\Box a^\nu & = \partial_\mu p^{\nu\mu} - \frac{1}{3} \partial^\nu p
		\\
		& = \partial_\mu p^{\nu\mu} + \partial^\nu \partial_\mu a^\mu \vphantom{\frac{1}{3}} \,. \nonumber
	\end{align}
These equations result in
	\begin{align}
		\Box a^\nu - \partial^\nu (\partial_\mu a^\mu) & = j^\nu + 2\, q^2\, \bar{\phi}\, \phi\, a^\nu \,, \label{eq:waveequationphoton}
	\end{align}
which is exactly the source equation for electromagnetic waves in scalar electrodynamics.

\subsection{Klein-Gordon-Maxwell system - in the Lagrangian picture}
\label{sec:kgmaxwelllagrange}

In the last chapter we showed that the Hamilton density (\ref{eq:hamiltonianfinal}) (with complex Klein-Gordon matter fields), which was derived via canonical transformation theory, leads to the well known equations for scalar electrodynamics, e.g. Eqs.\ (\ref{eq:kgmaxwelleq}), (\ref{eq:kgmaxwelleq2}) and (\ref{eq:waveequationphoton}). In this chapter we prove the consistency of this derivation. Therefore we Legendre-transform Hamilton density (\ref{eq:hamiltonfinalkg}) to find the appropriate Lagrange density. Subsequently we calculate the Euler-Lagrange equation (\ref{eq:eulerlagrangeequation}) for $\bar{\phi}$, $\phi$ and $a_\mu$.
\\
\\
The Legendre-transformation of Eq.\ (\ref{eq:hamiltonfinalkg}) reads
	\begin{align}
		\mathcal{L} & = \bar{\pi}^\mu (\partial_\mu \phi) + (\partial_\mu \bar{\phi}) \pi^\mu + p^{\nu\mu} (\partial_\mu a_\nu) - \mathcal{H} \label{eq:lagrangianfinal}
		\\
		& = \left[ (\partial_\mu \bar{\phi}) + iq\, \bar{\phi}\, a_\mu \right] \left[ (\partial^\mu \phi) - iq\, \phi\, a^\mu \right] - m^2 \bar{\phi}\, \phi + \frac{1}{2} (\partial_\mu a_\nu) (\partial_\mu a_\nu) - \frac{1}{2} (\partial_\mu a^\mu) (\partial_\mu a^\mu) \,, \nonumber
	\end{align}
where we used Eqs.\ (\ref{eq:legbarpi}), (\ref{eq:legpi}), (\ref{eq:gaugecantwo}) and (\ref{eq:gaugecantwotrace}). This is exactly the Maxwell-Lagrangian\footnote{Usually the Maxwell-Lagrangian is expressed via the field strength tensor $f_{\mu\nu} \equiv \partial_\mu a_\nu - \partial_\nu a_\mu$. Up to surface terms this representation coincides with the one given here, see \cite{greiner96}}.
\\
\\
We continue by calculating the Euler-Lagrange equations for $\bar{\phi}$ and $\phi$,
	\begin{align}
		0 & = \partial_\mu \frac{\partial \mathcal{L}}{\partial (\partial_\mu \bar{\phi})} - \frac{\partial \mathcal{L}}{\partial \bar{\phi}}
		\\
		& = (\partial_\mu - iq\, a_\mu) \left[ (\partial^\mu \phi) - iq\, \phi\, a^\mu \right] + m^2 \phi \vphantom{\frac{\partial \mathcal{L}}{\partial (\partial_\mu \bar{\phi})}} \nonumber
		\\
		& = (\mathcal{D}_\mu \mathcal{D}^\mu + m^2)\, \phi \vphantom{\frac{\partial \mathcal{L}}{\partial (\partial_\mu \bar{\phi})}} \,, \nonumber
		\\
		0 & = \partial_\mu \frac{\partial \mathcal{L}}{\partial (\partial_\mu \phi)} - \frac{\partial \mathcal{L}}{\partial \phi}
		\\
		& = (\partial_\mu + iq\, a_\mu) \left[ (\partial^\mu \bar{\phi}) + iq\, \bar{\phi}\, a^\mu \right] + m^2 \bar{\phi} \vphantom{\frac{\partial \mathcal{L}}{\partial (\partial_\mu \bar{\phi})}} \nonumber
		\\
		& = (\mathcal{D}^{*}_\mu \mathcal{D}^{* \mu} + m^2)\, \bar{\phi} \vphantom{\frac{\partial \mathcal{L}}{\partial (\partial_\mu \bar{\phi})}} \,. \nonumber
	\end{align}
Both coincide with the equations of motion (\ref{eq:kgmaxwelleq}) and (\ref{eq:kgmaxwelleq2}) found in the last chapter.
\\
\\
In our last step we compute the Euler-Lagrange equation (\ref{eq:eulerlagrangeequation}) for $a_\mu$,
	\begin{align}
		0 & = \partial_\mu \frac{\partial \mathcal{L}}{\partial (\partial_\mu a_\nu)} - \frac{\partial \mathcal{L}}{\partial a_\nu}
		\\
		& = \partial_\mu [ (\partial^\mu a^\nu) - \eta^{\mu\nu} (\partial_\alpha a^\alpha) ] - j^\nu - 2\, q^2\, \bar{\phi}\, \phi\, a^\nu \vphantom{\frac{\partial \mathcal{L}}{\partial (\partial_\mu a_\nu)}} \nonumber
		\\
		& = [ \Box a^\nu - \partial^\nu (\partial_\mu a^\mu)  ] - j^\nu - 2\, q^2\, \bar{\phi}\, \phi\, a^\nu \vphantom{\frac{\partial \mathcal{L}}{\partial (\partial_\mu a_\nu)}} \,. \nonumber
	\end{align}
This result coincides with (\ref{eq:waveequationphoton}), which completes the consistency check.

\section{Outlook}
\label{sec:outlook}
Gauge theory is the basis of modern quantum field theory. Except for the gravitational interaction all fundamental forces (electromagnetism, the strong and the weak interaction) are commonly formulated as gauge theories in a Lagrangian picture \cite{tHooft94,weinberg96}.
\\
The present paper demonstrates how gauge theories can be naturally formalized in a covariant Hamilton formalism, via exemplifying the procedure for the simplest non-trivial example, scalar electrodynamics ($U(1)$-symmetry). Covariant canonical transformations provide a guideline for the gauging procedure and globally ensure that the principle of least action is maintained. Starting with a complex massive field and the global symmetry the formalism directly ``proposes'' minimal coupling\footnote{Note, that even in this formal derivation of gauge theory minimal coupling is only sufficient, but not necessary.}. Sticking to this new canonical formalism the transformation behaviour, as well as dynamical terms in the Hamiltonian/Lagrangian, of the gauge field, are natural and unavoidable consequences. Klein-Gordon-Maxwell equations follow directly.
\\
\\
Nonetheless there are still open questions:
\\
The artificial choice of the integration constant $y^{\nu\mu}$ (\ref{eq:problemgleichung}) determines the transformation behaviour of the canonically conjugate gauge field which in turn determines the form of the gauge field's kinetic term in the Hamiltonian and the corresponding Lagrangian. We believe that identifying  this link with help of the canonical transformation formalism\footnote{This includes the application of Noether's theorem in the Hamilton formalism. Work along these lines is in progress.} is a step towards a better understanding of the remaining freedom in the formulation of gauge field dynamics, and that further analysis of the option for choosing the integration constant $y^{\nu\mu}$ in (\ref{eq:problemgleichung}) will lead to a stringent derivation of gauge theory with four single assumptions - Covariance/Poincar\'e invariance in flat space-time, the principle of least action (including a first order Lagrangian field theory), global $U(1)$-invariance of the matter field and minimal coupling - and the demand for local $U(1)$-invariance of the resulting theory.
\\
\\
Despite this last problem it appears natural to formulate all gauge theories in this framework, since it provides a clear and structured track without loopholes and ambiguities. Generalizations to higher $SU(N)$-symmetries are possible and parts of the calculation are already sketched in \cite{struckmeier08}: Local phase transformations (\ref{eq:transformationphi}) and (\ref{eq:transformationphibar}) are substituted by local unitary matrices. The calculation is similar yet more lengthy due to non-commutating objects. The generalization of the underlying massive matter to Dirac, Proca, Rarita-Schwinger, or even higher spin fields \cite{greiner96,koenigstein15} does not change the formalism, since only the underlying global symmetry determines the actual gauging procedure.
\\
\\
Further applications of this formalism, which are already under progress, could be spontaneous breaking of symmetries via inhomogeneous transformations, the standard model in the covariant Hamilton framework, or even a gauge theory of general relativity  (compare with first approaches in \cite{StrRei12,struckmeier13}). Another aftermath of this formalism could be covariant quantisation via generalised Poisson brackets.
\\
\\
\textbf{Acknowledgements:} The authors thank D.~D.~Dietrich for inspiring and very valuable discussions. The authors further thank E.~I.~Guendelman, S.~Hofmann and Y.~Verbin for useful discussions. We acknowledge support through the Helmholtz Graduate School HIRe for FAIR, through the Hessian Excellence initiative LOEWE HIC for FAIR, and by the Carl Wilhelm Fueck foundation, Frankfurt am Main.

\bibliography{extLagHam}
\bibliographystyle{plain}

\appendix

\addcontentsline{toc}{section}{Appendices}
\section{First Appendix}

\subsection{Euler-Lagrange equation}
\label{app:euler-lagrangeequation}
In this appendix we provide a short derivation of the Euler-Lagrange equation (\ref{eq:eulerlagrangeequation}) from the variational principle. The starting point will be the minimization of the action $\mathcal{S}$, which is given by
	\begin{align}
		\mathcal{S} & = \int_{R} \mathrm{d}^4 x\, \mathcal{L} (\phi,\, \partial_\alpha \phi,\, x) \,. \label{eq:action}
	\end{align}
We claim that classical physical systems will always evolve among to stationary configurations on the ``path'' with minimal action. Therefore we claim that the variation of the field $\delta\phi$ has to vanish on the space-time surface $\partial R$ and calculate
	\begin{align}
		0 & \overset{!}{=} \int_{R} \mathrm{d}^4 x\, \delta \mathcal{L} (\phi,\, \partial_\alpha \phi,\, x) \label{eq:variationsrechnung}
		\\
		& = \int_{R} \mathrm{d}^4 x\, \left[ \frac{\partial \mathcal{L}}{\partial \phi} \delta \phi + \frac{\partial \mathcal{L}}{\partial (\partial_\mu \phi)} \delta (\partial_\mu \phi) \right] \,. \nonumber
	\end{align}
At this point we use the infinitesimal variation of the field $\phi$,
	\begin{align}
		\varphi & = \phi + \delta \phi \,,
	\end{align}
to calculate the variation of the derivative of the field,

	\begin{align}
		\delta (\partial_\mu \phi) & = \partial_\mu \varphi - \partial_\mu \phi \label{eq:variationableitung}
		\\
		& = \partial_\mu (\delta \phi) \,. \nonumber
	\end{align}
We find that variation and differentiation commute. This trick is used to modify Eq.\ (\ref{eq:variationsrechnung}),
	\begin{align}
		0 & \overset{!}{=} \int_{R} \mathrm{d}^4 x\, \left[ \frac{\partial \mathcal{L}}{\partial \phi} \delta \phi + \frac{\partial \mathcal{L}}{\partial (\partial_\mu \phi)} \partial_\mu (\delta \phi) \right] \,.
	\end{align}
Integrating by parts yields
	\begin{align}
		0 & \overset{!}{=} \int_{R} \mathrm{d}^4 x\, \left\{ \left[ \frac{\partial \mathcal{L}}{\partial \phi} - \partial_\mu \frac{\partial \mathcal{L}}{\partial (\partial_\mu \phi)} \right] \delta\phi + \partial_\mu \left[ \frac{\partial \mathcal{L}}{\partial (\partial_\mu \phi)} \delta \phi \right]  \right\} \label{eq:herleitungeulerlagrange}
		\\
		& = \int_{R} \mathrm{d}^4 x\, \left[ \frac{\partial \mathcal{L}}{\partial \phi} - \partial_\mu \frac{\partial \mathcal{L}}{\partial (\partial_\mu \phi)} \right] \delta\phi + \cancel{\int_{\partial R} \mathrm{d}^4 S_\mu\, \frac{\partial \mathcal{L}}{\partial (\partial_\mu \phi)} \delta \phi} \nonumber
		\\
		& = \int_{R} \mathrm{d}^4 x\, \left[ \frac{\partial \mathcal{L}}{\partial \phi} - \partial_\mu \frac{\partial \mathcal{L}}{\partial (\partial_\mu \phi)} \right] \delta\phi \,, \nonumber
	\end{align}
where we converted the total divergence into a surface integral using the divergence theorem. Finally we were allowed to neglect this term since we claimed that the variation of the field $\delta\phi$ has to vanish on the surface $\partial R$.
In consequence the last line of Eq.\ (\ref{eq:herleitungeulerlagrange}) yields Eq.\ (\ref{eq:eulerlagrangeequation}).

\subsection{Surface terms}
\label{app:surfaceterms}
This appendix provides a formal proof of the invariance of the Euler-Lagrange equation (\ref{eq:eulerlagrangeequation}) under transformation (\ref{eq:surfaceterms}).\\
We want to show that $\mathcal{L}^\prime$ fulfils the Euler-Lagrange equation (\ref{eq:eulerlagrangeequation}), e.g.
	\begin{align}
		0 & = \partial_\mu \frac{\partial \mathcal{L}^\prime}{\partial (\partial_\mu \phi)} - \frac{\partial \mathcal{L}^\prime}{\partial \phi}
	\end{align}
Therefore we have to insert Eq.\ (\ref{eq:surfaceterms}) in the expression on the r.h.s.. We claim that $\mathcal{L}$ fulfils the Euler-Lagrange equation, which means that we only have to show that same applies for the surface term,
	\begin{align}
		\partial_\mu \frac{\partial \mathcal{L}^\prime}{\partial (\partial_\mu \phi)} - \frac{\partial \mathcal{L}^\prime}{\partial \phi} & = \partial_\mu \frac{\partial}{\partial (\partial_\mu \phi)} \left( \frac{\partial \mathcal{F}^\alpha}{\partial \phi} \partial_\alpha \phi + \left. \partial_\alpha \mathcal{F}^\alpha \right|_{expl} \right) - \frac{\partial}{\partial \phi} \left( \frac{\partial \mathcal{F}^\alpha}{\partial \phi} \partial_\alpha \phi + \left. \partial_\alpha \mathcal{F}^\alpha \right|_{expl} \right)
		\\
		& = \partial_\mu \frac{\partial \mathcal{F}^\mu}{\partial \phi} - \frac{\partial^2 \mathcal{F}^\alpha}{\partial \phi\, \partial \phi} \partial_\alpha \phi - \left. \frac{\partial^2 \mathcal{F}^\alpha}{\partial \phi\, \partial x^\alpha} \right|_{expl} \vphantom{\left. \frac{\partial^2 \mathcal{F}^\alpha}{\partial \phi\, \partial x^\alpha} \right|_{expl}} \nonumber
		\\
		& =  \frac{\partial^2 \mathcal{F}^\alpha}{\partial \phi\, \partial \phi} \partial_\alpha \phi + \left. \frac{\partial^2 \mathcal{F}^\alpha}{\partial \phi\, \partial x^\alpha} \right|_{expl} - \frac{\partial^2 \mathcal{F}^\alpha}{\partial \phi\, \partial \phi} \partial_\alpha \phi - \left. \frac{\partial^2 \mathcal{F}^\alpha}{\partial \phi\, \partial x^\alpha} \right|_{expl} \nonumber
		\\
		& = 0 \vphantom{\left. \frac{\partial^2 \mathcal{F}^\alpha}{\partial \phi\, \partial x^\alpha} \right|_{expl}} \,. \nonumber
	\end{align}	
This completes the proof.

\subsection{Point transformations}
\label{app:pointtransformations}
In this appendix we prove the invariance of the Euler-Lagrange equation (\ref{eq:eulerlagrangeequation}) under point transformations. We claim that $\mathcal{L}$ fulfils the Euler-Lagrange equation (\ref{eq:eulerlagrangeequation}). Next we point transform the field $\phi$
	\begin{align}
		\phi & = \phi (\Phi,\, x)
	\end{align}
and insert it into $\mathcal{L}$ which yields a transformed Lagrange density
	\begin{align}
		\tilde{\mathcal{L}}\left( \Phi,\, \partial_\beta \Phi,\, x \right) & = \mathcal{L} \left[\phi(\Phi,\, x),\, \partial_\alpha \phi(\Phi,\, \partial_\beta \Phi,\, x),\, x \right] \,.
	\end{align}
A useful relation will be
	\begin{align}
		\partial_\mu \phi & = \frac{\partial \phi}{\partial \Phi} \partial_\mu \Phi + \left. \partial_\mu \phi \right|_{expl} \longrightarrow \frac{\partial (\partial_\mu \phi)}{\partial (\partial_\nu \Phi)} = \delta^\mu_\nu \frac{\partial \phi}{\partial \Phi} \,. \label{eq:pointtransuseful}
	\end{align}
Now we are prepared to calculate
	\begin{align}
		\partial_\mu \frac{\partial \tilde{\mathcal{L}}}{\partial (\partial_\mu \Phi)} - \frac{\partial \tilde{\mathcal{L}}}{\partial \Phi} & = \partial_\mu \left[ \frac{\partial \mathcal{L}}{\partial (\partial_\nu \phi)} \frac{\partial (\partial_\nu \phi)}{\partial (\partial_\mu \Phi)} \right] - \left[ \frac{\partial \mathcal{L}}{\partial \phi} \frac{\partial \phi}{\partial \Phi} + \frac{\partial \mathcal{L}}{\partial (\partial_\nu \phi)} \frac{\partial (\partial_\nu \phi)}{\partial \Phi} \right]
		\\
		& = \partial_\mu \left[ \frac{\partial \mathcal{L}}{\partial (\partial_\mu \phi)} \frac{\partial \phi}{\partial \Phi} \right] - \left[ \frac{\partial \mathcal{L}}{\partial \phi} \frac{\partial \phi}{\partial \Phi} + \frac{\partial \mathcal{L}}{\partial (\partial_\nu \phi)} \frac{\partial (\partial_\nu \phi)}{\partial \Phi} \right] \vphantom{\frac{\partial \tilde{\mathcal{L}}}{\partial (\partial_\mu \Phi)}} \nonumber
		\\
		& = \left[ \partial_\mu \frac{\partial \mathcal{L}}{\partial (\partial_\mu \phi)} \right] \frac{\partial \phi}{\partial \Phi}  + \frac{\partial \mathcal{L}}{\partial (\partial_\mu \phi)} \left( \partial_\mu \frac{\partial \phi}{\partial \Phi} \right) - \left[ \frac{\partial \mathcal{L}}{\partial \phi} \frac{\partial \phi}{\partial \Phi} + \frac{\partial \mathcal{L}}{\partial (\partial_\nu \phi)} \frac{\partial (\partial_\nu \phi)}{\partial \Phi} \right] \vphantom{\frac{\partial \tilde{\mathcal{L}}}{\partial (\partial_\mu \Phi)}} \nonumber
		\\
		& = \left[ \partial_\mu \frac{\partial \mathcal{L}}{\partial (\partial_\mu \phi)} - \frac{\partial \mathcal{L}}{\partial \phi} \right] \frac{\partial \phi}{\partial \Phi} + \frac{\partial \mathcal{L}}{\partial (\partial_\nu \phi)} \left[ \frac{\partial (\partial_\nu \phi)}{\partial \Phi} - \frac{\partial (\partial_\nu \phi)}{\partial \Phi} \right] \vphantom{\frac{\partial \tilde{\mathcal{L}}}{\partial (\partial_\mu \Phi)}} \nonumber
		\\
		& = 0 \,. \vphantom{\frac{\partial \tilde{\mathcal{L}}}{\partial (\partial_\mu \Phi)}} \nonumber
	\end{align}
Finally we showed that $\tilde{\mathcal{L}}$ fulfils the Euler-Lagrange equation (\ref{eq:eulerlagrangeequation}) as well.

\subsection{Canonical equations}
\label{app:canonicalequations}
In this section we derive the canonical equations (\ref{eq:canonicalone}) and (\ref{eq:canonicaltwo}) via the variational principle. In analogy to Sec.\ \ref{app:euler-lagrangeequation} we start with the action integral (\ref{eq:action}) and express the Lagrange density $\mathcal{L}$ immediately by its Legendre transformation (\ref{eq:legendretransformation}),
	\begin{align}
		\mathcal{S} & = \int_{R} \mathrm{d}^4 x\, \left[ \pi^\mu \partial_\mu \phi - \mathcal{H}(\phi, \pi^\alpha, x) \right] \,. \label{eq:hamiltonaction}
	\end{align}
We claim that the variation of the action has to vanish and find,
	\begin{align}
		0 & \overset{!}{=} \int_{R} \mathrm{d}^4 x\, \delta \left[ \pi^\mu \partial_\mu \phi - \mathcal{H}(\phi, \pi^\alpha, x) \right]
		\\
		& = \int_{R} \mathrm{d}^4 x\, \left[ \partial_\mu \phi\, \delta\pi^\mu + \pi^\mu \delta (\partial_\mu \phi) - \frac{\partial \mathcal{H}}{\partial \phi} \delta \phi - \frac{\partial \mathcal{H}}{\partial \pi^\mu} \delta \pi^\mu \right] \,. \nonumber
	\end{align}
To continue we make use of relation (\ref{eq:variationableitung}) to exchange the derivative and the variation in the second term. Subsequent we partially integrate the same expression an neglect the surface term, since the variation of the field $\delta\phi$ vanishes on the space-time surface $\partial R$,
	\begin{align}
		0 & \overset{!}{=} \int_{R} \mathrm{d}^4 x\, \left[ \partial_\mu \phi\, \delta\pi^\mu + \pi^\mu \partial_\mu (\delta \phi) - \frac{\partial \mathcal{H}}{\partial \phi} \delta \phi - \frac{\partial \mathcal{H}}{\partial \pi^\mu} \delta \pi^\mu \right]
		\\
		& = \int_{R} \mathrm{d}^4 x\, \left[ \partial_\mu \phi\, \delta\pi^\mu - (\partial_\mu \pi^\mu) \delta \phi + \partial_\mu (\pi^\mu \delta \phi) - \frac{\partial \mathcal{H}}{\partial \phi} \delta \phi - \frac{\partial \mathcal{H}}{\partial \pi^\mu} \delta \pi^\mu \right] \nonumber
		\\
		& = \int_{R} \mathrm{d}^4 x\, \left[ \partial_\mu \phi - \frac{\partial \mathcal{H}}{\partial \pi^\mu} \right] \delta\pi^\mu - \int_{R} \mathrm{d}^4 x\, \left[ \partial_\mu \pi^\mu + \frac{\partial \mathcal{H}}{\partial \phi} \right] \delta \phi + \cancel{\int_{\partial R} \mathrm{d}^4 S_\mu\, \pi^\mu \delta \phi}  \nonumber
		\\
		& = \int_{R} \mathrm{d}^4 x\, \left[ \partial_\mu \phi - \frac{\partial \mathcal{H}}{\partial \pi^\mu} \right] \delta\pi^\mu - \int_{R} \mathrm{d}^4 x\, \left[ \partial_\mu \pi^\mu + \frac{\partial \mathcal{H}}{\partial \phi} \right] \delta \phi \nonumber \,.
	\end{align}
The variation of the field $\delta \phi$ and the variation of the canonically conjugate fields $\delta \pi^\mu$ are independent and therefore both integrands have to vanish separately.

\subsection{Surface terms in the Hamilton density}
\label{app:hamiltonsurface terms}
In this section we show that surface terms, which leave the Euler-Lagrange equation invariant, see App.\ \ref{app:surfaceterms}, correspond to additional partial space-time derivatives of the same vector functions $\mathcal{F}^\mu (\phi,\, x)$ in Hamilton densities, which leave the canonical equations (\ref{eq:canonicalone}) and (\ref{eq:canonicaltwo}) invariant. We start with Eq.\ (\ref{eq:surfaceterms}) and define associated fields and associated canonically conjugate momentum fields,
	\begin{align}
		\phi^\prime & \equiv \phi \,, \vphantom{\frac{\partial (\partial_\alpha \mathcal{F}^\alpha)}{\partial (\partial_\nu \phi)}}
		\\
		\pi^{\prime\mu} & \equiv \frac{\partial \mathcal{L}^\prime }{\partial (\partial_\mu \phi^\prime)}
		\\
		& = \left[ \frac{\partial \mathcal{L}}{\partial (\partial_\nu \phi)} + \frac{\partial (\partial_\alpha \mathcal{F}^\alpha)}{\partial (\partial_\nu \phi)} \right] \frac{\partial (\partial_\nu \phi)}{\partial (\partial_\mu \phi^\prime)} \nonumber
		\\
		& = \pi^\mu + \frac{\partial}{\partial (\partial_\mu \phi)} \left( \frac{\partial \mathcal{F}^\alpha}{\partial \phi} \partial_\alpha \phi + \left. \partial_\alpha \mathcal{F}^\alpha \right|_{expl} \right) \nonumber
		\\
		& = \pi^\mu + \frac{\partial \mathcal{F}^\mu}{\partial \phi} \,. \nonumber
	\end{align}
We use the Legendre transformation (\ref{eq:legendretransformation}) for $\mathcal{L}^\prime$ to find the Hamilton density $\mathcal{H}^\prime$ and its relation to $\mathcal{H}$,
	\begin{align}
		\mathcal{H}^\prime & = \pi^{\prime\mu} \partial_\mu \phi^\prime - \mathcal{L}^\prime \vphantom{\left( \pi^\mu + \frac{\partial \mathcal{F}^\mu}{\partial \phi} \right)} \vphantom{\frac{\partial (\partial_\mu \phi^\prime)}{\partial (\partial_\nu \phi)}}
		\\
		& = \left( \pi^\mu + \frac{\partial \mathcal{F}^\mu}{\partial \phi} \right) \frac{\partial (\partial_\mu \phi^\prime)}{\partial (\partial_\nu \phi)} \partial_\nu \phi - \mathcal{L} - \partial_\mu \mathcal{F}^\mu \nonumber
		\\
		& = \mathcal{H} + \frac{\partial \mathcal{F}^\mu}{\partial \phi} \partial_\mu \phi - \left( \frac{\partial \mathcal{F}^\mu}{\partial \phi} \partial_\mu \phi + \left. \partial_\mu \mathcal{F}^\mu \right|_{expl} \right) \vphantom{\frac{\partial (\partial_\mu \phi^\prime)}{\partial (\partial_\nu \phi)}} \nonumber
		\\
		& = \mathcal{H} - \left. \partial_\mu \mathcal{F}^\mu \right|_{expl} \,. \vphantom{\frac{\partial (\partial_\mu \phi^\prime)}{\partial (\partial_\nu \phi)}} \nonumber
	\end{align}
To complete the proof we simply calculate the canonical equations (\ref{eq:canonicalone}) and (\ref{eq:canonicaltwo}) for $\mathcal{H}^\prime$. We use that the canonical equations hold for $\mathcal{H}$,
	\begin{align}
		\frac{\partial \mathcal{H}^\prime}{\partial \phi^\prime} & = \frac{\partial \mathcal{H}}{\partial \phi} \frac{\partial \phi}{\partial \phi^\prime} + \frac{\partial \mathcal{H}}{\partial \pi^\mu} \frac{\partial \pi^\mu}{\partial \phi} \frac{\partial \phi}{\partial \phi^\prime} - \left. \frac{\partial^2 \mathcal{F}^\mu}{\partial \phi\,  \partial x^\mu}  \right|_{expl} \frac{\partial \phi}{\partial \phi^\prime}
		\\
		& = - \partial_\mu \pi^\mu - \partial_\mu \phi \frac{\partial^2 \mathcal{F}^\mu}{\partial \phi\, \partial \phi} - \left. \frac{\partial^2 \mathcal{F}^\mu}{\partial \phi\,  \partial x^\mu}  \right|_{expl} \nonumber
		\\
		& = - \partial_\mu \pi^\mu - \partial_\mu \left( \frac{\partial}{\partial \phi} \mathcal{F}^\mu \right) \nonumber
		\\
		& = - \partial_\mu \pi^{\prime\mu} \,. \vphantom{\frac{\partial^2 \mathcal{F}^\mu}{\partial \phi\, \partial \phi}} \nonumber
	\end{align}
We find that the first canonical equation is also fulfilled for $\mathcal{H}^\prime$ whereas the second canonical equation is calculated via
	\begin{align}
		\frac{\partial \mathcal{H}}{\partial \pi^{\prime\mu}} & = \frac{\partial \mathcal{H}}{\partial \pi^\nu} \frac{\partial \pi^\nu}{\partial \pi^{\prime\mu}}
		\\
		& = \partial_\mu \phi \vphantom{\frac{\partial \mathcal{H}}{\partial \pi^{\prime\mu}}} \nonumber
		\\
		& = \partial_\mu \phi^\prime \vphantom{\frac{\partial \mathcal{H}}{\partial \pi^{\prime\mu}}} \nonumber
	\end{align}
and is also valid for $\mathcal{H}^\prime$.

\subsection{Point transformations for Hamilton densities}
\label{app:pointtransformationsham}
In this appendix we show that point transformations defined in Eq.\ (\ref{eq:pointtransformationsforfields}) leave the dynamics and therefore the canonical equations (\ref{eq:canonicalone}) and (\ref{eq:canonicaltwo}) invariant. We start with the calculation of the canonically conjugate field for a point transformation,
	\begin{align}
		\Pi^\mu & \equiv \frac{\partial \tilde{\mathcal{L}}}{\partial (\partial_\mu \Phi)}
		\\
		& = \frac{\partial \mathcal{L}}{\partial (\partial_\nu \phi)} \frac{\partial (\partial_\nu \phi)}{\partial (\partial_\mu \Phi)} \nonumber
		\\
		& = \pi^\mu \frac{\partial \phi}{\partial \Phi} \,, \nonumber
	\end{align}
where we used Eq.\ (\ref{eq:pointtransuseful}) in the last line. The Hamilton density $\tilde{\mathcal{H}}$ is defined by
	\begin{align}
		\tilde{\mathcal{H}} & \equiv \Pi^\mu \partial_\mu \Phi - \tilde{\mathcal{L}} \vphantom{\frac{\partial \Phi}{\partial \phi}}
		\\
		& = \pi^\mu \frac{\partial \phi}{\partial \Phi} \left( \frac{\partial \Phi}{\partial \phi} \partial_\mu \phi + \left. \partial_\mu \Phi \right|_{expl} \right) -\mathcal{L} \nonumber
		\\
		& = \mathcal{H} + \Pi^\mu \left. \partial_\mu \Phi \right|_{expl} \,, \vphantom{\frac{\partial \Phi}{\partial \phi}} \nonumber
	\end{align}
where we assumed that $\mathcal{H}$ is the Legendre transform of $\mathcal{L}$, compare (\ref{eq:legendretransformation}). Now we are ready to show that $\tilde{\mathcal{H}}$ fulfils the canonical equations (\ref{eq:canonicalone}) and (\ref{eq:canonicaltwo}) if $\mathcal{H}$ fulfils the corresponding ones.
	\begin{align}
		\frac{\partial \tilde{\mathcal{H}}}{\partial \Phi} + \partial_\mu \Pi^\mu & = \frac{\partial \mathcal{H}}{\partial \phi} \frac{\partial \phi}{\partial \Phi} + \frac{\partial \mathcal{H}}{\partial \pi^\mu} \cancel{\frac{\partial \pi^\mu}{\partial \Phi}} + \cancel{\frac{\partial}{\partial \Phi} \left( \Pi^\mu \left. \partial_\mu \Phi \right|_{expl} \right)} + \cancel{\frac{\partial \Pi^\mu}{\partial \phi}} \partial_\mu \phi + \frac{\partial \Pi^\mu}{\partial \pi^\nu} \partial_\mu \pi^\nu
		\\
		& = - \partial_\mu \pi^\mu \frac{\partial \phi}{\partial \Phi} + \partial_\mu \pi^\mu \frac{\partial \phi}{\partial \Phi} \vphantom{\cancel{\frac{\partial \Pi^\mu}{\partial \phi}}} \nonumber
		\\
		& = 0 \,. \vphantom{\cancel{\frac{\partial \Pi^\mu}{\partial \phi}}} \nonumber
	\end{align}
Therefore the first equation is proven.
	\begin{align}
		\frac{\partial \tilde{\mathcal{H}}}{\partial \Pi^\mu} - \partial_\mu \Phi & = \frac{\partial \mathcal{H}}{\partial \phi} \cancel{\frac{\partial \phi}{\partial \Pi^\mu}} + \frac{\partial \mathcal{H}}{\partial \pi^\nu} \frac{\partial \pi^\nu}{\partial \Pi^\mu} + \frac{\partial}{\partial \Pi^\mu} \left( \Pi^\nu \left. \partial_\nu \Phi \right|_{expl} \right) - \frac{\partial \Phi}{\partial \phi} \partial_\mu \phi - \left. \partial_\mu \Phi \right|_{expl}
		\\
		& = \frac{\partial \Phi}{\partial \phi} \partial_\mu \phi + \left. \partial_\mu \Phi \right|_{expl} - \frac{\partial \Phi}{\partial \phi} \partial_\mu \phi - \left. \partial_\mu \Phi \right|_{expl} \vphantom{\cancel{\frac{\partial \phi}{\partial \Pi^\mu}}} \nonumber
		\\
		& = 0\,. \vphantom{\cancel{\frac{\partial \phi}{\partial \Pi^\mu}}} \nonumber
	\end{align}

\section{Second Appendix}
	
\subsection{Symmetry relations for the matter field}
\label{app:proofcanonicaltrans}

In this appendix we will list and calculate all symmetry relations for the transformation rules (\ref{eq:transphi}), (\ref{eq:transphibar}), (\ref{eq:transmomentum}) and (\ref{eq:transmomentumbar}). In contrast to real scalar fields, see Sec.\ \ref{sec:canonicaltransformations}, there are six symmetry relations per generating function.
\\
\\
Generalising the symmetry relation of $f_1^\mu$, given in Eq.\ (\ref{eq:f1symm}), to complex fields we find
	\begin{align}
		\frac{\partial^2 c^\mu}{\partial \bar{\phi}\, \partial \phi} = \frac{\partial \pi^\mu}{\partial \phi} & = \frac{\partial^2 f_1^\mu}{\partial \bar{\phi}\, \partial \phi} = \frac{\partial \bar{\pi}^\mu}{\partial \bar{\phi}} = \frac{\partial^2 c^\mu}{\partial \bar{\phi}\, \partial \phi} \,,
		\\
		0 = \frac{\partial \bar{\pi}^\mu}{\partial \Phi} & = \frac{\partial^2 f_1^\mu}{\partial \phi\, \partial \Phi} = - \frac{\partial \bar{\Pi}^\mu}{\partial \phi} = \frac{\partial^2 c^\mu}{\partial \phi\, \partial \phi} e^{i\Lambda} \,,
		\\
		0 = \frac{\partial \Pi^\mu}{\partial \Phi} & = - \frac{\partial^2 f_1^\mu}{\partial \bar{\Phi}\, \partial \Phi} = \frac{\partial \bar{\Pi}^\mu}{\partial \bar{\Phi}} = 0 \,,
		\\
		0 = \frac{\partial \pi^\mu}{\partial \bar{\Phi}} & = \frac{\partial^2 f_1^\mu}{\partial \bar{\phi}\, \partial \bar{\Phi}} = - \frac{\partial \Pi^\mu}{\partial \bar{\phi}}  = \frac{\partial^2 c^\mu}{\partial \bar{\phi}\, \partial \bar{\phi}} e^{-i\Lambda} \,,
		\\
		0 = \frac{\partial \pi^\mu}{\partial \Phi} & = \frac{\partial^2 f_1^\mu}{\partial \bar{\phi}\, \partial \Phi} = - \frac{\partial \bar{\Pi}^\mu}{\partial \bar{\phi}} = \frac{\partial^2 c^\mu}{\partial \phi\, \partial \bar{\phi}} e^{i\Lambda} \,,
		\\
		0 = \frac{\partial \bar{\pi}^\mu}{\partial \bar{\Phi}} & = \frac{\partial^2 f_1^\mu}{\partial \phi\, \partial \bar{\Phi}} = - \frac{\partial \Pi^\mu}{\partial \phi}  = \frac{\partial^2 c^\mu}{\partial \bar{\phi}\, \partial \phi} e^{-i\Lambda} \,.
	\end{align}
For symmetry relation (\ref{eq:f2symm}) of $f_2^\mu$ the generalisation yields
	\begin{align}
		\frac{\partial^2 c^\mu}{\partial \bar{\phi}\, \partial \phi} = \frac{\partial \pi^\mu}{\partial \phi} & = \frac{\partial^2 f_2^\mu}{\partial \phi\, \partial \bar{\phi}} = \frac{\partial \bar{\pi}^\mu}{\partial \bar{\phi}} = \frac{\partial^2 c^\mu}{\partial \bar{\phi}\, \partial \phi} \,,
		\\
		0 = \frac{\partial \bar{\pi}^\mu}{\partial \Pi^\nu} & = \frac{\partial^2 f_2^\mu}{\partial \phi\, \partial \Pi^\nu} = \delta^\mu_\nu \frac{\partial \bar{\Phi}}{\partial \phi} = 0 \,,
		\\
		0 = \delta^\mu_\nu \frac{\partial \Phi}{\partial \Pi^\alpha} & = \frac{\partial^2 f_2^\mu}{\partial \bar{\Pi}^\nu\, \partial \Pi^\alpha} = \delta^\mu_\alpha \frac{\partial \bar{\Phi}}{\partial \bar{\Pi}^\nu} = 0 \,,
		\\
		0 = \frac{\partial \pi^\mu}{\partial \bar{\Pi}^\nu} & = \frac{\partial^2 f_2^\mu}{\partial \bar{\phi}\, \partial \bar{\Pi}^\nu} = \delta^\mu_\nu \frac{\partial \Phi}{\partial \bar{\phi}} = 0 \,,
		\\
		\delta^\mu_\nu e^{i\Lambda} = \frac{\partial \pi^\mu}{\partial \Pi^\nu} & = \frac{\partial^2 f_2^\mu}{\partial \bar{\phi}\, \partial \Pi^\nu} = \delta^\mu_\nu \frac{\partial \bar{\Phi}}{\partial \bar{\phi}} = \delta^\mu_\nu e^{i\Lambda} \,,
		\\
		\delta^\mu_\nu e^{-i\Lambda} = \frac{\partial \bar{\pi}^\mu}{\partial \bar{\Pi}^\nu} & = \frac{\partial^2 f_2^\mu}{\partial \phi\, \partial \bar{\Pi}^\nu} = \delta^\mu_\nu \frac{\partial \Phi}{\partial \phi} = \delta^\mu_\nu e^{-i\Lambda} \,.
	\end{align}
Generalising the symmetry relation (\ref{eq:f3symm}) of $f_3^\mu$ to complex scalar fields we find
	\begin{align}
		0 = \delta^\mu_\nu \frac{\partial \phi}{\partial \pi^\alpha} & = - \frac{\partial^2 f_3^\mu}{\partial \bar{\pi}^\nu\, \partial \pi^\alpha} = \delta^\mu_\alpha \frac{\partial \bar{\phi}}{\partial \bar{\pi}^\nu} = 0 \,,
		\\
		0 = \delta^\mu_\nu \frac{\partial \bar{\phi}}{\partial \Phi} & = - \frac{\partial^2 f_3^\mu}{\partial \pi^\nu\, \partial \Phi} = \frac{\partial \bar{\Pi}^\mu}{\partial \pi^\nu} = 0 \,,
		\\
		0 = \frac{\partial \Pi^\mu}{\partial \Phi} & = - \frac{\partial^2 f_3^\mu}{\partial \bar{\Phi}\, \partial \Phi} = \frac{\partial \bar{\Pi}^\mu}{\partial \bar{\Phi}} = 0 \,,
		\\
		0 = \delta^\mu_\nu \frac{\partial \phi}{\partial \bar{\Phi}} & = - \frac{\partial^2 f_3^\mu}{\partial \bar{\pi}^\nu\, \partial \bar{\Phi}} = \frac{\partial \Pi^\mu}{\partial \bar{\pi}^\nu} = 0 \,,
		\\
		\delta^\mu_\nu e^{i\Lambda} = \delta^\mu_\nu \frac{\partial \phi}{\partial \Phi} & = - \frac{\partial^2 f_3^\mu}{\partial \bar{\pi}^\nu\, \partial \Phi} = \frac{\partial \bar{\Pi}^\mu}{\partial \bar{\pi}^\nu} = \delta^\mu_\nu e^{i\Lambda} \,,
		\\
		\delta^\mu_\nu e^{-i\Lambda} = \delta^\mu_\nu \frac{\partial \bar{\phi}}{\partial \bar{\Phi}} & = - \frac{\partial^2 f_3^\mu}{\partial \pi^\nu\, \partial \bar{\Phi}} = \frac{\partial \Pi^\mu}{\partial \pi^\nu} = \delta^\mu_\nu e^{-i\Lambda} \,.
	\end{align}
The last generalisation of (\ref{eq:f4symm}) corresponds to generating function $f_4^\mu$,
	\begin{align}
		0 = \delta^\mu_\alpha \frac{\partial \phi}{\partial \pi^\beta} & = - \frac{\partial^2 f_4^\mu}{\partial \bar{\pi}^\alpha\, \partial \pi^\beta} = \delta^\mu_\beta \frac{\partial \bar{\phi}}{\partial \bar{\pi}^\alpha} = 0 \,,
		\\
		0 = \delta^\mu_\alpha \frac{\partial \bar{\phi}}{\partial \Pi^\beta} & = - \frac{\partial^2 f_4^\mu}{\partial \pi^\alpha\, \partial \Pi^\beta} = - \delta^\mu_\beta \frac{\partial \bar{\Phi}}{\partial \pi^\alpha} = 0 \,,
		\\
		0 = \delta^\mu_\alpha \frac{\partial \Phi}{\partial \Pi^\beta} & = \frac{\partial^2 f_4^\mu}{\partial \bar{\Pi}^\alpha\, \partial \Pi^\beta} = \delta^\mu_\beta \frac{\partial \bar{\Phi}}{\partial \bar{\Pi}^\alpha} = 0 \,,
		\\
		0 = \delta^\mu_\alpha \frac{\partial \phi}{\partial \bar{\Pi}^\beta} & = - \frac{\partial^2 f_4^\mu}{\partial \bar{\pi}^\alpha\, \partial \bar{\Pi}^\beta} = - \delta^\mu_\beta \frac{\partial \Phi}{\partial \bar{\pi}^\alpha} = 0 \,,
		\\
		0 = \delta^\mu_\alpha \frac{\partial \phi}{\partial \Pi^\beta} & = - \frac{\partial^2 f_4^\mu}{\partial \bar{\pi}^\alpha\, \partial \Pi^\beta} = - \delta^\mu_\beta \frac{\partial \bar{\Phi}}{\partial \bar{\pi}^\alpha} = 0 \,,
		\\
		0 = \delta^\mu_\alpha \frac{\partial \bar{\phi}}{\partial \bar{\Pi}^\beta} & = - \frac{\partial^2 f_4^\mu}{\partial \pi^\alpha\, \partial \bar{\Pi}^\beta} = - \delta^\mu_\beta \frac{\partial \Phi}{\partial \pi^\alpha} = 0 \,.
	\end{align}
	
\subsection{Symmetry relations for the gauge field}
\label{app:symmetryforgauge}

In this appendix we calculate the symmetry relations for the transformations rules (\ref{eq:transgauge}) and (\ref{eq:transkonjgauge}) of the gauge field. Symmetry relations among the gauge field and the matter fields are neglected since all of them vanish.
\\
\\
The four symmetry relations (\ref{eq:f1symm}), (\ref{eq:f2symm}), (\ref{eq:f3symm}) and (\ref{eq:f4symm}) for gauge fields read
	\begin{align}
		0 & = \frac{\partial p^{\nu\mu}}{\partial A_\alpha}  = \frac{\partial^2 f_1^\mu}{\partial a_\nu\, \partial A_\alpha} = - \frac{\partial P^{\alpha\mu}}{\partial a_\nu} = - \frac{\partial^2 c^{\prime\mu}}{\partial a_\alpha\, \partial a_\nu} \,,
		\\
		\delta^\mu_\beta \delta^\nu_\alpha & = \frac{\partial p^{\nu\mu}}{\partial P^{\alpha\beta}} = \frac{\partial^2 f_2^\mu}{\partial a_\nu\, \partial P^{\alpha\beta}} = \delta^\mu_\beta \frac{\partial A_\alpha}{a_\nu} = \delta^\mu_\beta \delta^\nu_\alpha \,,
		\\
		\delta^\mu_\beta \delta^\nu_\alpha & = \delta^\mu_\beta \frac{\partial a_\alpha}{\partial A_\nu} = - \frac{\partial^2 f_3^\mu}{\partial p^{\alpha\beta}\, \partial A_\nu} = \frac{\partial P^{\nu\mu}}{\partial p^{\alpha\beta}} = \delta^\mu_\beta \delta^\nu_\alpha \,,
		\\
		0 & = - \delta^\mu_\beta \frac{\partial a_\alpha}{\partial P^{\sigma\rho}} = \frac{\partial^2 f_4^\mu}{\partial p^{\alpha\beta}\, \partial P^{\sigma\rho}} = \delta^{\mu\rho} \frac{\partial A_\sigma}{\partial p^{\alpha\beta}} = 0 \,.
	\end{align}
	
\subsection{The integration constant}
\label{app:integrationconstant}
In this appendix we give a brief motivation why the integration constant $y^{\nu\mu}$ might be not arbitrary, although we do not find a clear restriction without further assumptions about the finalized theory.
We know from Eq.\ (\ref{eq:kleinpzugrossp}) that $y^{\nu\mu}$ affects the transformation rule for the canonically conjugate field. However, if we make further assumptions about the resulting theory, $y^{\nu\mu}$ cannot be chosen arbitrarily any more, since it is not independent from the transformation behaviour (\ref{eq:transgaugefield}) of the gauge field $a_\mu$, due to the fact that $p^{\nu\mu}$ and $\partial_\mu a_\nu$ are connected via
	\begin{align}
		p^{\nu\mu} & \equiv \frac{\partial \mathcal{L}_a}{\partial (\partial_\mu a_\nu)} \,.
	\end{align}
Now we argue that if the formalism leads to a dynamical Lagrangian for the gauge field $a_\mu$ (which is the fact) than this dynamical Lagrangian should be somehow quadratic in the derivatives of $a_\mu$. This is a an assumption and therefore a clear restriction in the generality of the whole theory! In consequence
	\begin{align}
		p^{\nu\mu} & = \frac{\partial \mathcal{L}_a}{\partial (\partial_\mu a_\nu)} \propto (\partial^\bullet a^\bullet) \label{eq:bulleteq}
	\end{align}
with dummy indices ``$\bullet$'', which means that $p^{\nu\mu}$ is in some way proportional to any first derivatives of $a_\mu$. With Eqs.\ (\ref{eq:transgaugefield}) and (\ref{eq:kleinpzugrossp}) it follows for the transformed canonically conjugate field that,
	\begin{align}
		P^{\nu\mu} + y^{\nu\mu} & \propto \partial^\bullet \left[ A^\bullet + \frac{1}{q} (\partial^\bullet \Lambda) \right] \,.
	\end{align}
According for the fact that the final Lagrangian $\mathcal{L}_a$ and consistently Eq.\ (\ref{eq:bulleteq}) have to be form-invariant, we find
	\begin{align}
		y^{\nu\mu} & \propto \partial^\bullet \partial^\bullet \Lambda \,.
	\end{align}
Therefore the remaining most general choice for $y^{\nu\mu}$ is
	\begin{align}
		y^{\nu\mu} & = t_1 \eta^{\nu\mu} (\eta_{\alpha\beta} \partial^\alpha \partial^\beta \Lambda) + t_2 (\partial^\nu \partial^\mu \Lambda)
		\\
		& = t_1 \eta^{\nu\mu} (\Box \Lambda) + t_2 (\partial^\nu \partial^\mu \Lambda) \nonumber
	\end{align}
with yet independent constants $t_1$ and $t_2$.\\
Choosing $t_1 = - t_2$ we find that the integration constant $y^{\nu\mu}$ reduces to a projection operator orthogonal to the derivative of $\Lambda$. In fact this means that the inhomogeneity $\partial_\mu \Lambda$ in the transformation behaviour of the gauge field $a_\mu$, see Eq.\ (\ref{eq:transgaugefield}), is exactly orthogonal to the inhomogeneous term $y^{\nu\mu}$ in the transformation behaviour of the canonically conjugate gauge fields, see Eq.\ (\ref{eq:kleinpzugrossp}). We believe that this cannot be by accident and might be a guidepost towards the solution of the remaining problem.
	
\subsection{The Klein-Gordon field in the Lagrangian and Hamilton picture}
\label{app:kleingordonfield}

In this appendix we give a brief discussion of the complex Klein-Gordon field in the Lagrangian and covariant Hamiltonian picture, compare \cite{greiner96}. In general the Lagrange density for a complex field reads
	\begin{align}
		\mathcal{L} (\phi,\, \bar{\phi},\, \partial_\alpha \phi,\, \partial_\alpha \bar{\phi}) & = (\partial_\mu \bar{\phi}) (\partial^\mu \phi) - m^2 \bar{\phi}\, \phi \,. \label{eq:kleingordonlagrangian}
	\end{align}
Inserting this Lagrange density in the Euler-Lagrange equation (\ref{eq:eulerlagrangeequation}) for the complex conjugate field $\bar{\phi}$ as well as the field $\phi$ this yields
	\begin{align}
		(\Box + m^2)\, \phi & = 0 \,, \label{eq:kgequation}
		\\
		(\Box + m^2)\, \bar{\phi} & = 0 \,, \label{eq:kgequationcomplex}
	\end{align}
which are the corresponding equations of motion, in fact the Klein-Gordon equations.
\\
\\
In the next step we show that the same result is also obtained in the Hamilton picture. Therefore we start with Legendre-transforming Eq.\ (\ref{eq:kleingordonlagrangian}). To do so we have to calculate the canonical conjugate fields of $\bar{\phi}$ and $\phi$ first, compare Eq.\ (\ref{eq:canonicallyconjugatefield}),
	\begin{align}
		\pi^\mu & = \frac{\partial \mathcal{L}}{\partial (\partial_\mu \bar{\phi})} = \partial^\mu \phi \,,
		\\
		\bar{\pi}^\mu & = \frac{\partial \mathcal{L}}{\partial (\partial_\mu \phi)} = \partial^\mu \bar{\phi} \,.
	\end{align}
Subsequent using the Legendre transformation (\ref{eq:legendretransformation}) the Hamilton density reads then,
	\begin{align}
		\mathcal{H} (\phi,\, \bar{\phi},\, \pi^\alpha,\, \bar{\pi}^\alpha) & = \bar{\pi}^\mu (\partial_\mu \phi) + (\partial_\mu \bar{\phi}) \pi^\mu - \mathcal{L} \label{eq:kleingordonhamiltonian}
		\\
		& = \bar{\pi}_\mu \pi^\mu + m^2 \bar{\phi}\, \phi \,. \nonumber
	\end{align}
We calculate the four canonical equations, compare Eq.\ (\ref{eq:canonicalone}) and (\ref{eq:canonicaltwo}),
	\begin{align}
		- \partial_\mu \bar{\pi}^\mu & = \frac{\partial \mathcal{H}}{\partial \phi} = m^2 \bar{\phi} \,, \vphantom{\frac{\partial \mathcal{H}}{\partial \bar{\phi}}} \label{eq:kgone}
		\\
		\partial_\mu \bar{\phi} & = \frac{\partial \mathcal{H}}{\partial \pi^\mu} = \bar{\pi}_\mu \,, \vphantom{\frac{\partial \mathcal{H}}{\partial \bar{\phi}}} \label{eq:kgtwo}
		\\
		- \partial_\mu \pi^\mu & = \frac{\partial \mathcal{H}}{\partial \bar{\phi}} = m^2 \phi \,, \vphantom{\frac{\partial \mathcal{H}}{\partial \bar{\phi}}} \label{eq:kgthree}
		\\
		\partial_\mu \phi & = \frac{\partial \mathcal{H}}{\partial \bar{\pi}^\mu} = \pi_\mu \,, \vphantom{\frac{\partial \mathcal{H}}{\partial \bar{\phi}}} \label{eq:kgfour}
	\end{align} 
and insert Eq.\ (\ref{eq:kgtwo}) in Eq.\ (\ref{eq:kgone}) as well as Eq.\ (\ref{eq:kgfour}) in Eq.\ (\ref{eq:kgthree}), which leads to the Klein-Gordon equations (\ref{eq:kgequation}) and (\ref{eq:kgequationcomplex}). Therefore the covariant Hamilton and the Lagrangian picture are completely equivalent.
\\
\\
As a consistency check we could Legendre back-transform the Klein-Gordon Hamilton density (\ref{eq:kleingordonhamiltonian}) to the Klein-Gordon Lagrange density (\ref{eq:kleingordonlagrangian}) under the use of Eq.\ (\ref{eq:kgtwo}) and (\ref{eq:kgfour}). This calculation is trivial and therefore left to the reader.

\end{document}